\documentclass[12pt,a4paper]{article}

\usepackage{amsmath}
\usepackage{amssymb}
\usepackage{graphicx}
\usepackage{bm}
\def\one{\mbox{\bf 1}}

\def\half{\frac{1}{2}}
\def\eps{\epsilon}
\def\RR{\mathbb R}
\def\NN{\mathbb N}
\def\CC{\mathbb C}
\def\PP{\mathbb P}
\def\EE{\mathbb E}

\def\qed{\hfill \vrule height6pt width4pt depth2pt}

\newcommand{\non}{\nonumber}
\newcommand{\Tr}{\mathop{\rm Tr}\,}

\newcommand{\ul}{\underline}
\newcommand{\ol}{\overline}

\newcommand{\Pf}{\noindent\textbf{Proof. }}

\newcommand{\be}{\begin{equation}}
\newcommand{\ee}{\end{equation}}
\newtheorem{theorem}{Theorem}[section]
\newtheorem{lemma}{Lemma}[section]
\newtheorem{prop}{Proposition}[section]

\newtheorem{cor}{Corollary}[section]

\begin{document}

\title{Quantum Large Deviations}
\author{T. C. Dorlas\footnote{Dublin Institute for Advanced Studies, School of Theoretical Physics, 10 Burlington road, Dublin 04, Ireland.} }
%\hfill DIAS-24-28 
\maketitle

\begin{abstract}
We reconsider the quantum analogue of Varadhan's Theorem proved by Petz, Raggio and Verbeure\cite{PRV1989}. They proved
this theorem using standard techniques in quantum statistical mechanics of lattice systems to arrive at a variational formula over states on a C$^*$ algebra, which can subsequently be reduced to a variational formula in terms of a single real variable. 
In this paper a new proof is given using a quantum version of the large deviation analysis together with
the Trotter product formula. The proof is subsequently extended to the general case of $q$ non-commuting variables resulting in a variational formula for general mean-field quantum spin systems as first  derived by Raggio and Werner~\cite{RW1989}.
\end{abstract}

Data statement: No datasets were generated or analysed.

\section{The PRV theorem}

Petz, Raggio and Verbeure \cite{PRV1989} proved a quantum version of Varadhan's Theorem \cite{Var1966}. Their theorem is stated in terms of general C$^*$ algebras, but here we consider
only the case of a product of finite-dimensional algebras. 
Let $\cal M$ be the algebra of all complex $m \times m$ matrices and let $H$ and $X$ be 
Hermitian matrices  $H, X \in {\cal M}$. 
Consider the tensor product algebras ${\cal M}_n = {\cal M}^{\otimes n}$. 
We denote by $X^{(n)}$ the element of ${\cal M}_n$ given by
$$ X^{(n)} = \frac{1}{n}(X_1 + \dots + X_n), $$
where $X_k = \one \otimes \dots \otimes X \otimes \dots \otimes \one$ is a copy of $X$ in the $k$-th factor of ${\cal M}_n$.  The PRV theorem then states the following.

\begin{theorem}[Petz-Raggio-Verbeure] \label{PRVThm}
If $f:[-||X||,||X||] \to \RR$ is a continuous function then
\begin{equation} \label{PRV}  \lim_{n \to \infty} \frac{1}{n} \ln \Tr e^{n[f(X^{(n)}) - H^{(n)}]} = \sup_{u \in [-||X||,||X||]} [f(u) - I(u)] + \ln \Tr(e^{-H}), \end{equation}  
where $I:[-||X||,||X||] \to [0,+\infty]$ is the Legendre transform of
\begin{equation} \label{Qgenfn} C(s) = \ln \Tr e^{-H + s X} - \ln \Tr e^{-H}. \end{equation}
\end{theorem}

The proof uses C$^*$-algebra techniques. The theorem was generalized by Raggio and Werner \cite{RW1989} to general quantum mean-field spin systems.

A special case of Theorem~\ref{PRVThm} was proved in \cite{Dor2009} using the Donsker-Varadhan Theorem \cite{DV3} and the Trotter-product formula. Here we prove the full theorem also using 
the Trotter product formula and large-deviation techniques for the upper bound. 

First we make some observations about the cumulant generating function 
$C(s)$ given by (\ref{Qgenfn}).
%$$ C(t) = \ln \frac{\Tr(e^{tX - H})}{\Tr(e^{-H})}. $$

\begin{lemma} \label{QLD-L1-1} The cumulant generating function $C(t)$ is convex. 
Its derivative is given by
\begin{equation} \label{Cprime} C'(t) = \frac{\Tr(X\,e^{tX-H})}{\Tr(e^{tX-H})}. \end{equation} 
\end{lemma} 

\Pf We compute the first and second derivatives using the Duhamel formula:
$$ e^{A + B} = e^A + \int_0^1 e^{s(A+B)} B\, e^{(1-s) A} ds. $$
(This is derived by differentiating $e^{s(A+B)} e^{-sA}$ with respect to $s$.) 
Setting $A=tX-H$ and $B=uX$, we have
\begin{eqnarray*} e^{(t+u)X - H} &=&  e^A + 
u \int_0^1 e^{s(uX + A)} X\,e^{(1-s)A} ds \\ &=& e^A + 
u \int_0^1 e^{sA} X\, e^{(1-s)A} ds + u \int_0^1 (e^{s(uX + A)} - e^{sA})\,X\,e^{(1-s)A)} ds.
\end{eqnarray*}
Since the integral in the second term tends to 0 as $u \to 0$, we find that 
\begin{eqnarray} \frac{d}{dt} e^{tX-H} &=& \frac{d}{du}\big|_{u=0} e^{uX + A} = 
\int_0^1 e^{sA}\,X\,e^{(1-s)A} ds \non \\ &=& \int_0^1 e^{s(tX-H)}\,X\, e^{(1-s)(tX - H)} ds. 
\end{eqnarray}
Taking the trace, we obtain (\ref{Cprime}).
Differentiating again, we have, writing $Z(t) = \Tr(e^{tX-H})$,
\begin{eqnarray*} C''(t) &=& \frac{1}{Z(t)} \int_0^1 
\Tr \left[ X\,e^{s(tX-H)}\,X\, e^{(1-s)(tX - H)} \right]\, ds -
\left(\frac{\Tr[X\,e^{tX-H}]}{Z(t)} \right)^2 \\ &=& 
\frac{1}{Z(t)} \int_0^1 \Tr \left[ (X-\EE_t(X))\,e^{s(tX-H)}\,(X-\EE_t(X))\, e^{(1-s)(tX - H)}
\right]\, ds, \end{eqnarray*}
where 
\begin{equation} \EE_t(X) = \frac{1}{Z(t)}{\Tr[X\,e^{tX-H})} = C'(t). \end{equation}
The expression 
\begin{equation} \langle A\,|\,B \rangle_{\rm Bog} = 
\frac{1}{Z(t)} \int_0^1 \Tr \left[ A^*\,e^{s(tX-H)}\,B\, e^{(1-s)(tX - H)} \right]\, ds
\end{equation} is called the \textbf{Bogoliubov scalar product.} It is easily shown to be a 
scalar product. We can thus write
\begin{equation} C''(t) = \langle (X-\EE_t(X))\,|\,(X-\EE_t(X)) \rangle_{\rm Bog} \geq 0. \end{equation}
This means in particular that the cumulant generating function $C$ is convex.  \qed

\begin{cor} \label{QLD-cor1} 
Let $S$ be the convex hull of the spectrum of $X$, $S = {\rm co}(\sigma(X))$, i.e. 
$S = [\lambda_-, \lambda_+]$, where $\lambda_-$
and $\lambda_+$ are the smallest and largest eigenvalues of $X$. 
Let $I$ be the Legendre transform of $C$. Then $I(u) < +\infty$ if and only if  $u \in S$. 
Moreover, $\lim_{u \uparrow \lambda_+} I'(u) = +\infty$ and 
$\lim_{u \downarrow \lambda_-} I'(u) = -\infty$. \end{cor}

\Pf Since $X \leq \lambda_+ \one$, where $\one \in {\cal M}$ is the identity matrix, we have,
noting that $\Tr(X\,e^{tX-H}) = \Tr [e^{(tX-H)/2} X e^{(tX-H)/2}]$, $C'(t) \leq \lambda_+$.
Similarly, $C'(t) \geq \lambda_-$ for all $t \in \RR$. Moreover, since $C'(t)$ is increasing,
we have $\lim_{t \to \pm\infty} C'(t) = \lambda_\pm$. Taking $t \to \pm \infty$ in 
$I(u) = \sup_{t \in \RR} [tu - C(t)]$, we see that $I(u) = +\infty$ if $u > \lambda_+$ or $u < \lambda_-$.
For $u \in (\lambda_-,\lambda_+)$, we have $I(u) = t(u) u - C(t(u))$, where $t(u)$ is 
given by $u = \EE_{t(u)}(X) = C'(t(u))$. Then $I'(u) = t(u) \to \pm\infty$ as $u \to \lambda_\pm$.
By perturbation theory, we have in fact that $C(t) \sim \lambda_\pm t + \Tr (P_\pm H)$ as $t \to \pm \infty$, where  $P_\pm$ is the projection onto the eigenspace corresponding to the eigenvalues $\lambda_\pm$ of $X$. 
This implies that $I(\lambda_\pm) <+\infty$. \qed

\setcounter{equation}{0}

\section{The associated quantum stochastic process}

The proof of the PRV theorem is divided into an upper bound and a lower bound. 
Our proof of the upper bound is similar to the large deviation upper bound, but for a sequence of
complex path measures. 

We now prove the existence of a \textit{Quantum Stochastic Process} (QSP), i.e. a complex-valued
measure on paths representing the above trace.

\begin{theorem} \label{QSPThm}Suppose that $X$ and $H$ are self-adjoint $m \times m$ matrices.
Let $S = \sigma(X)$ be the spectrum of $X$. 
There exists a complex-valued bounded Radon measure $\kappa$ on the Skorokhod space
$D([0,1],S)$ such that for any finite partition $0 \leq t_1 < \dots < t_N \leq 1$, and Borel
subsets $A_1, \dots, A_N \subset S$,
\begin{align} &\kappa(\xi(t_i) \in A_i\,(i=1,\dots,N)) = \non \\ & \qquad = 
\Tr \left[e^{-(1-t_N) H} P_{A_N} e^{-(t_N-t_{N-1}) H} \dots P_{A_1} e^{-t_1 H} \right],
\label{kappadef} \end{align}
where $P_A$ is the spectral projection of $X$ corresponding to the set $A$.
\end{theorem}

\Pf This is similar to the existence of a Feynman integral on finite sets: see \cite{Thomas1996} and also \cite{DT2009}. We diagonalize $X$ and write $H$ as a matrix in the corresponding basis. As in \cite{Dor2009}, we first adjust the diagonal of $H$. Defining the 
diagonal matrix $H_D$ with diagonal matrix elements
\begin{equation} \label{HD} H_D(k) = H_{k,k} - \sum_{j \neq k} |H_{j,k}| \end{equation}
 ($k \in \NN_m = \{1,\dots,m\}$), set 
 \begin{equation} \tilde{H} = H - H_D, \end{equation}
so that in particular,
\begin{equation} \tilde{H}_{k,k} = \sum_{j\neq k} |H_{j,k}|. \end{equation}

We first define measures on the set of all paths $\NN_m^{[0,1]}$ (with product topology)
with values in $\NN_m = \{1,\dots,m\}$. Given a subdivision 
$\sigma:\,0 \leq t_1 < \dots < t_N \leq 1$, and subsets $A_1, \dots, A_N \subset \NN_m$, define
\begin{eqnarray} \label{musigma} \mu^\sigma (A_1 \times \dots \times A_N)  &=&
\sum_{k_1 \in A_1} \dots \sum_{k_N \in A_N} \left(e^{-(1-t_N+t_1) \tilde{H}} \right)_{k_1,k_N} 
\\ && \qquad \times 
\left(e^{-(t_N-t_{N-1}) \tilde{H}} \right)_{k_N,k_{N-1}} \dots \left(e^{-(t_2-t_1) \tilde{H}} \right)_{k_2,k_1}  \non \end{eqnarray} 
This defines a complex-valued measure $\mu^\sigma$ on $\NN_m^\sigma$. 
It is obvious that these measures form a projective system 
in the sense that if $\sigma'$ is a refinement of $\sigma$ then the restriction of 
$\mu^{\sigma'}$ to the functions $\Phi$
depending only on the points of $\sigma$ equals $\mu^\sigma$:
\begin{equation} \int \Phi \circ \pi_{\sigma',\sigma} \, d\mu^{\sigma'} = \int \Phi\,d\mu^\sigma, \end{equation} for $\Phi \in {\cal C}(\NN_m^\sigma)$. 
($\pi_{\sigma',\sigma}(\xi)$ is the restriction of $\xi: \sigma' \to \NN_m$ to $\sigma$.)

We now introduce the positivity-preserving operator (matrix) $Q_t$ with kernel
\begin{equation} Q_t(i,j) = |(e^{-t \tilde{H}})_{i,j}|. \end{equation}
Since $(e^{-(t+s) \tilde{H}})_{i,j} = \sum_{k=1}^m (e^{-t \tilde{H}})_{i,k} (e^{-s \tilde{H}})_{k,j}$,
it follows that
\begin{equation} Q_{t+s}(i,j)  \leq \sum_{k=1}^m Q_t(i,k) Q_s(k,j). \label{Qincrease} \end{equation}
We argue that this implies that
\begin{equation} ||Q_{t+s} || \leq ||Q_t||\,||Q_s||. \end{equation}
Indeed, if $A$ and $B$ are symmetric positivity-preserving matrices and $A_{i,j} \leq B_{i,j}$ for all $i,j$ then $||A|| \leq ||B||$.
For, by the Perron-Frobenius theorem, the eigenvector $v$ of $A$ with maximal eigenvalue $||A||$ has non-negative components, and hence
$$ (Bv)_i = \sum_j B_{i,j} v_j \geq \sum_j A_{i,j} v_j = ||A||\,v_i $$
and $$ ||B|| = \sup_{u:\,||u||=1} \langle u, Bu\rangle \geq \langle v\ Bv\rangle \geq ||A||, $$
assuming that $v$ is normalized.

We need an upper bound on $||Q_t||$. For small $t$, we can write
\begin{equation} \label{approxprop} 
(e^{-t \tilde{H}})_{i,j} = \delta_{i,j} - t \tilde{H}_{i,j} + O(t^2) \end{equation}  
and therefore
\begin{eqnarray*} ||Q_t|| &=& \sup_{u:\,||u|| = 1} \left\| \sum_j Q_t(i,j) u_j \right\| \\
&\leq & \sup_{u:\,||u||=1} \left\| u + t \sum_j |\tilde{H}_{i,j}|\,|u_j| + O(t^2) \right\| 
\leq 1 + t ||R|| + O(t^2), \end{eqnarray*}
where $R$ denotes the matrix with matrix elements $|\tilde{H}_{i,j}|$.
Subdividing $[0,t]$ into $p$ small intervals, we have
$$ ||Q_t|| \leq \bigg(1+ \frac{t}{p} ||R|| + O(t^2/p^2)\bigg)^p $$ and taking the limit $p \to \infty$,
\begin{equation} ||Q_t|| \leq e^{t\,||R||}, \end{equation}

For symmetric positivity-preserving matrices $Q$, the matrix elements $Q_{i,j}$ are bounded by
$||Q||$ because if we take $v_i = \delta_{i,i_0}$ then $Q_{i_0,i_0} = \langle v,Qv \rangle$ and if
$v_i = \frac{1}{\sqrt{2}} (\delta_{i,i_1} \pm \delta_{i,i_2})$ then $\langle v,Qv \rangle =
\half(Q_{i_1,i_1} + Q_{i_2,i_2} \pm 2 Q_{i_1,i_2}) \leq ||Q||$. Therefore,
\begin{equation} \label{Qbound} Q_t(i,j) \leq e^{t\,||R||}. \end{equation}

Given a subdivision $\sigma = \{t_1, \dots, t_N\}$ of $[0,1]$, i.e.
$0\leq t_1 < \dots < t_N < 1$, the variation of the measure $\mu^\sigma$ is given by
\begin{equation} \int \Phi\,d|\mu^\sigma| = \sum_{i_1,\dots,i_N \in \NN_m} \Phi(i_1,\dots,i_N)\,
|\mu^\sigma| (\{(i_1,\dots,i_N)\}), \end{equation}

\begin{eqnarray} |\mu^\sigma| (\{(i_1,\dots,i_N)\}) &=& |\mu^\sigma (\{(i_1,\dots,i_N)\})| \non \\
&=& Q_{t_N-t_{N-1}}(i_N,i_{N-1}) \dots Q_{t_2-t_1}(i_2,i_1) Q_{1-t_N+t_1}(i_1,i_{N}). \non \\ && \label{modmusigma} \end{eqnarray}
Note that the inequality (\ref{Qincrease}) implies that the right-hand side is increasing in
$\sigma$. Moreover, it is bounded above by (\ref{Qbound}) and therefore converges as $\sigma$
gets finer. If $\Phi \in {\cal C}(\NN_m^\sigma)$  then
\begin{align} &\left| \int \Phi\, d|\mu^\sigma| \right| \non \\ &= \left| \sum_{i_1,\dots, i_N
\in \NN_m} \Phi(i_1,\dots,i_N)\, Q_{t_N-t_{N-1}}(i_N,i_{N-1}) \dots
Q_{1-t_N+t_1}(i_1,i_N) \right| \non \\ &\leq  ||\Phi||_\infty \Tr \left[Q_{t_N-t_{N-1}} \dots
Q_{1-t_N+t_1}\right] \non \\ &\leq  ||\Phi||_\infty ||Q_{t_N-t_{N-1}}||\dots
||Q_{t_2-t_1}||\,\Tr[Q_{1-t_N+t_1}] \non \\ &\leq  ||\Phi||_\infty \prod_{k=2}^N ||Q_{t_k-t_{k-1}}|| \,
\Tr[Q_{1-t_N+t_1}] \leq m\,e^{||R||} ||\Phi||_\infty.\end{align} It follows that
\begin{equation} \int \Phi^\sigma\,d\mu = \lim_{\sigma'} \int \Phi^\sigma\,d\mu^{\sigma'} \end{equation}
exists and is bounded by $m\,e^{||R||} ||\Phi||_\infty$ for $\Phi^\sigma$ of the form $\Phi \circ \pi_\sigma$ with $\Phi \in {\cal C}(\NN_m^\sigma)$ where $\pi: \NN_m^{[0,1]} \to \NN_m^\sigma$ 
is the projection. Since these functions are dense in ${\cal C}(\NN_m^{[0,1]})$ the integral can
be extended to a continuous linear form on ${\cal C}(\NN_m^{[0,1]})$ and by the Riesz-Markov
theorem this defines a complex-valued Radon measure on $\NN_m^{[0,1]}$. Moreover, the 
measures $|\mu^\sigma|$ also converge to a measure $|\mu|$ on ${\cal C}(\NN_m^{[0,1]})$.

Note also that it follows from (\ref{approxprop}) that, for small $t > 0$,
\begin{equation} Q_t(i,j) = \begin{cases} 1 - t \,|\tilde{H}_{i,i}| + O(t^2) &\text{if $i=j$;} \\
t\,|\tilde{H}_{i,j}| + O(t^2) &\text{if $i \neq j$.} \end{cases} \end{equation}
Inserting this into (\ref{modmusigma}), we see that the limiting measure $|\mu|$ has the 
generating matrix $e^{t \Gamma}$, where 
\begin{equation} \Gamma_{i,j} = \begin{cases} -|\tilde{H}_{i,i}| = -\sum_{k \neq i} |H_{i,k}| 
&\text{if $i = j$;} \\ |H_{i,j}| &\text{if $i \neq j$.} \end{cases} \end{equation}
(This means that for a subdivision $\sigma$ of $[0,1]$, the image measure $\pi_\sigma(|\mu|)$
is given by 
\begin{eqnarray*} \lefteqn{\pi_\sigma(|\mu|)(\{i_1,\dots,i_N\})} \\ &=& 
e^{(t_N-t_{N-1}) \Gamma}(i_N,i_{N-1}) \dots e^{(t_2-t_1) \Gamma}(i_2,i_1)
e^{(1-t_N+t_1) \Gamma}(i_1,i_{N}). \end{eqnarray*}
Note that $\pi_\sigma(|\mu|) \neq |\mu^\sigma|$\ !)
The matrix $\Gamma$ is a Q-matrix (see \cite{Norris1997} or \cite{Doob1953}, Chapter VI, equation (1.6)),  which means that the transition matrix $e^{t \Gamma}$ determines a stationary 
random process $\xi(t)$ with values in $\NN_m$ such that 
$$ \PP(\xi(t)=k'\,|\,\xi(0) = k) = (e^{t \Gamma})_{k',k}. $$
The corresponding path measure with initial state $k_0 = k \in \NN_m$ is defined by 
$$ \nu_k(\xi(t_i) \in A_i\,(i=1,\dots,N)) = \sum_{k_1 \in A_1} \dots \sum_{k_N \in A_N} 
\prod_{i=1}^N \left( e^{(t_i-t_{i-1}) \Gamma} \right)_{k_i,k_{i-1}}. $$  % where $k_0=k$.
Then we have for $A \in {\cal B}(\NN_m^{[0,1]})$,
$$ |\mu|(A) =  \sum_{k \in \NN_m} \int_A \one_{\{\xi(1)=k\}} \nu_k(d\xi). $$

We now want to show that the measures $\mu$ and $|\mu|$ are concentrated on the Skorokhod space $D([0,1],\NN_m)$. 
For this we prove 

\begin{lemma} \label{QLD-L2-1} For given $\eta > 0$, there exists a compact set 
$K_\eta \subset D([0,1],\NN_m)$ such that 
$\pi_\sigma(|\mu|)(\pi^{-1}_\sigma(K_\eta)) \leq \eta$. 
Therefore $|\mu|$ and hence also $\mu$, is concentrated on $D([0,1],\NN_m)$.
\end{lemma}

\Pf
Recall (see \cite{Partha1967}, Chapter VII, Theorem 6.2) that a set $K \subset D([0,1])$ 
is compact if it is closed and bounded, 
and such that $\limsup_{\delta \downarrow 0}  \tilde{\omega}_\delta(\xi) = 0$ 
uniformly in $\xi \in K$, where $\tilde{\omega}$ is defined by 
\begin{eqnarray} \tilde{\omega}_\delta(\xi) &=& 
\max \left\{ \sup_{t-\delta/2 < t' \leq t \leq t'' < t+\delta/2}
(|\xi(t')-\xi(t)| \wedge |\xi(t'') - \xi(t)|), \right. \non \\ 
&&\qquad  \left. \sup_{0 \leq t < t+ \delta/2} |\xi(t) - \xi(0)|, 
\sup_{1-\delta/2 < t \leq 1} |\xi(t) - \xi(1)| \right\}. \end{eqnarray}
Defining 
$$ G_\delta = \{\xi \in D([0,1],\NN_m):\, \tilde{\omega}_\delta(\xi) \leq \eta\} $$ 
where $\eta < 1$,
we have that $\pi_\sigma^{-1}(i_1,\dots,i_N) \notin G_\delta$ if and only if there is at least one 
pair of jumps between unequal eigenvalues a distance less than $\delta$ apart. 
We subdivide $[0,1]$ into intervals of length $\delta$. If $\xi \notin G_\delta$ then there is a double interval of length $2\delta$ which contains points at distance at most $\delta$ at which $\xi$ has a jump. 
Consider such a double interval and let $t_{k_1}$ be the left-most point of $\sigma$ and $t_{k_2}$ the right-most point of $\sigma$ contained in this interval. Then the corresponding 
$|\mu|$-measure is bounded by 
\begin{eqnarray*} 
\lefteqn{\pi_\sigma(|\mu|)(\{\mbox{at least 2 jumps between}\,  t_{k_1}\,{\rm and}\, t_{k_2}\})}
\non \\ &\leq & \sum_{k=k_1}^{k_2-1} \sum_{k' = k+1}^{k_2}
\sum_{i, i_1,\dots,i_{k-1} \in \NN_m} \sum_{i_{k} \neq i_{k-1}} \sum_{i_{k'} \neq i_{k}} 
\sum_{i_{k'+1}, \dots, i_N \in \NN_m}  \\ && \times 
\left( e^{(1-t_N) \Gamma} \right)(i,i_{N})  \\ && \times
\left( e^{(t_{N}-t_{N-1}) \Gamma} \right)(i_{N},i_{N-1})  \dots  
\left( e^{(t_{k'+1}-t_{k'}) \Gamma} \right)(i_{k'+1},i_{k'})  \non \\ && \times 
\left( e^{(t_{k'}-t_{k'-1}) \Gamma} \right)(i_{k'},i_{k})  \\ && \times
\left( e^{(t_{k'-1}-t_{k'-2}) \Gamma} \right)(i_{k},i_{k})  \dots
\left( e^{(t_{k+1}-t_{k}) \Gamma} \right)(i_{k},i_{k})    \\ && \times
\left( e^{(t_{k}-t_{k-1}) \Gamma} \right)(i_{k},i_{k-1})  \\ && \times
\left( e^{(t_{k-1}-t_{k-2}) \Gamma} \right)(i_{k-1},i_{k-2}) \dots 
\left( e^{(t_2-t_1) \Gamma} \right)(i_{2},i_1) \left( e^{t_1 \Gamma} \right)(i_1,i).
\end{eqnarray*} 
This contracts to
\begin{eqnarray}   \label{2jumpprob} 
\lefteqn{\pi_\sigma(|\mu|)(\{\mbox{at least 2 jumps between}\,  t_{k_1}\,{\rm and}\, t_{k_2}\})}
\non \\ &\leq & 
\sum_{k=1}^{k_2-1} \sum_{k' = k+1}^{k_2}
\sum_{i,i_{k-1},i_{k'-1} \in \NN_m} \sum_{i_{k} \neq i_{k-1}} 
\sum_{i_{k'} \neq i_{k'-1}} 
\left( e^{(1-t_{k'}) \Gamma} \right)(i,i_{k'})  \non \\ && \times
\left( e^{(t_{k'}-t_{k'-1}) \Gamma} \right)(i_{k'},i_{k'-1}) 
\left( e^{(t_{k'-1}-t_{k}) \Gamma} \right)(i_{k'-1},i_{k}) \non \\ && \qquad \times
\left( e^{(t_{k}-t_{k-1}) \Gamma} \right)(i_{k},i_{k-1})  
\left( e^{t_{k-1} \Gamma} \right)(i_{k-1},i). \end{eqnarray}
Using the bound 
$$ || A(t)|| \leq t ||\Gamma||\, ||e^{t\Gamma}|| \mbox{ if } 
A(t)_{i,j} = (e^{t \Gamma})(i,j) (1-\delta_{i,j}), $$
we find that 
\begin{eqnarray} \lefteqn{\pi_\sigma(|\mu|)(\{\mbox{at least 2 jumps between}\, 
t_{k_1}\,{\rm and}\, t_{k_2}\})} \non \\ 
&\leq &  \sum_{k=k_1}^{k_2-1} \sum_{k' = k+1}^{k_2} (t_{k}-t_{k-1}) (t_{k'}-t_{k'-1})
||\Gamma||^2 ||e^{-(t_k-t_{k-1} + t_{k'}-t_{k'-1}) \Gamma}||\,\Tr(e^{\Gamma}) \non \\ &\leq &
4 C\,\delta^2 \end{eqnarray} for a constant $C$ since 
\begin{eqnarray*} \lefteqn{\sum_{k=k_1}^{k_2-1} \sum_{k'=k+1}^{k_2} 
(t_k-t_{k-1})(t_{k'} - t_{k'-1})} \\ &=&
\sum_{k=k_1}^{k_2-1} (t_k - t_{k-1})(t_{k_2} - t_{k}) 
\leq (t_{k_2-1}-t_{k_1-1})(t_{k_2}-t_{k_1}) < 4\delta^2. \end{eqnarray*}
Summing over the intervals it follows that 
\begin{equation} \label{typicalpaths} \pi_\sigma(|\mu|)(\pi^{-1}_\sigma(G_\delta)^c) \leq 4 C \delta, \end{equation}
unifiormly in $\sigma$.
Now taking $K_\delta = \bigcap_{n \in \NN} G_{\delta/n^2}$ we have that 
\begin{equation} \label{Kconc} 
|\mu|(K_\delta^c) \leq C \delta \sum_{n=1}^\infty \frac{1}{n^2} 
= \frac{2 \pi^2}{3} C \delta. \end{equation}
This proves that $|\mu|$ and therefore also $\mu$ is concentrated on $D([0,1],\NN_m)$. \qed

Integrating the function $\prod_{j=1}^N \phi_j$, where  
$\phi_j = e^{-(t_{j+1}-t_j) H_D} \one_{A_j}$  for $j=1,\dots,N$ with respect to the measure 
$\mu^\sigma$ given by (\ref{musigma}) we have
\begin{eqnarray}\int \prod_{i=1}^N \phi_j \, d\mu^\sigma &=&
\sum_{k_1 \in A_1} \dots \sum_{k_N \in A_N} \left(e^{-(1-t_N+t_1) \tilde{H}} \right)_{k_1,k_N} 
e^{-(1-t_N) H_D(k_N)}  \non \\
&& \qquad  \times \left(e^{-(t_N-t_{N-1}) \tilde{H}} \right)_{k_N,k_{N-1}} e^{-(t_N-t_{N-1}) H_D(k_{N-1})} \non \\ && \qquad \qquad 
\dots  \left(e^{-(t_2-t_1) \tilde{H}} \right)_{k_2,k_1} e^{-(t_2-t_1) H_D(k_1)}. \end{eqnarray} 
Fixing $t_1,\dots,t_N$ and the sets $A_1,\dots,A_N$, but refining the subdivision by 
adding additional points between $t_i$ and $t_{i+1}$, we obtain in the limit
\begin{eqnarray} \lefteqn{\Tr \left[ e^{-(1-t_N) (\tilde{H} + H_D)} \tilde{P}_{A_N} \dots 
e^{-(t_2-t_1)(\tilde{H} + H_D)} \tilde{P}_{A_1} e^{-t_1 \tilde{H}} \right] }  \non \\ &=& 
\int_{\pi_\sigma^{-1}(A_1 \times \dots \times A_N)} e^{-\int_0^1 H_D(\xi(t))\,dt} \, \mu(d\xi), 
\end{eqnarray}
where $\pi_\sigma(\xi) = (\xi(t_1), \dots,\xi(t_N))$, and $\tilde{P}_A$ is the projection onto the 
$X$-eigenspace corresponding to the eigenvalues $\lambda_i$ with $i \in A$. 
Let $\tilde{\lambda}: [1,m] \to S$ be the linear interpolation between the eigenvalues 
$\lambda_i$, i.e. $\tilde{\lambda}(x) = \lambda_i + x(\lambda_{i+1}-\lambda_i)$ if $x \in [1,m]$.
Then $\tilde{\lambda}(\tilde{P}_A) = P_{\tilde{\lambda}(A)}$. Defining the measure $\tilde{\mu}$ by 
\begin{equation} \tilde{\mu}(B) = \int_B e^{-\int_0^1 H_D(\xi(t))\,dt} \, \mu(d\xi) \end{equation} 
for Borel sets $B \subset D([0,1],\NN_m)$, and the image measure $\kappa$
by \begin{equation} \kappa = \tilde{\lambda}(\tilde{\mu}), \end{equation} 
it follows that (\ref{kappadef}) holds. \qed

Expanding  a function $F(X)$ into eigenprojections, we have
\begin{eqnarray} \lefteqn{\Tr \left[ e^{-(1-t_N) H} F(X) \dots 
e^{-(t_2-t_1) H} F(X) e^{-t_1 H} \right]} \non \\ &=&
\int_{\pi_\sigma^{-1}(A_1 \times \dots \times A_N)} 
F(\eta(t_1)) \dots, F(\eta(t_N))\,\kappa(d\eta),  \end{eqnarray}

By the above estimate (\ref{Kconc}), the limiting measure $|\mu|$ is also defined on $D([0,1],\NN_m)$.
Moreover, the typical paths have a finite number of jumps. The same therefore also holds for $\kappa$.
Therefore, if $f: S \to \RR$ is continuous and bounded, $f \circ \eta$ is Riemann integrable 
for almost all $\eta$ and 
$$ \frac{1}{N} \sum_{j=1}^N f(\eta(j/N)) \to \int_0^1 f(\eta(t))\,dt. $$
By the Lie-Trotter product formula, we therefore have the Feynman-Kac type formula
\begin{eqnarray} \Tr \left[e^{f(X) - H} \right] &=&  \lim_{N \to \infty} 
\Tr \left[\left( e^{-H/N} e^{f(X)/N} \right)^N \right] \non \\ &=& 
\lim_{N \to \infty} \int \exp \left[ \frac{1}{N} \sum_{j=1}^N f(\eta(j/N)) \right] \kappa(d\eta) \non \\ 
&=& \int e^{\int_0^1 f(\eta(t))\,dt} \kappa(d\eta).
\end{eqnarray}
Similarly, we have for the product measure
\begin{equation} \Tr \left[ e^{n(f(X^{(n)}) - H^{(n)})}] \right] = 
\int \exp \left[ n \int_0^1 f \left(\frac{1}{n} \sum_{k=1}^n \xi_k(t) \right) dt \right] 
\prod_{k=1}^n \kappa(d\xi_k). \end{equation} 
In the following we embed $D([0,1])$ into $L^2([0,1])$ and denote the image measures 
on $L^2([0,1])$ by the same symbols $\mu$ and $\kappa$. 
The Feynman-Kac integrals are then integrals over $L^2([0,1])$ and $L^2([0,1]^n)$
respectively.

\setcounter{equation}{0}

\section{The upper bound}

In order to prove the upper bound, we consider the product measure $|\mu|^{\otimes n}$ and  prove the following crucial lemma, which is due to Donsker and Varadhan~\cite{DV3}. We repeat their proof 
for completeness.

\begin{lemma} \label{QLD-L3-1} Given  $L > 0$, there exists a compact set $K_L \subset L^2([0,1])$
such that 
\begin{equation} \label{QLDtight} 
\limsup_{n \to \infty} \frac{1}{n} \ln |\mu|^{\otimes n} \bigg( \big\{ (\xi_k)_{k=1}^n:\,
\frac{1}{n} \sum_{k=1}^n \tilde{\lambda}(\xi_k(t)) \in K_L^c \big\} \bigg) < -L. 
\end{equation} 
\end{lemma}

\Pf  Let $\eps \in (0,\half]$, and choose $\delta \in (0,\eps)$ such that 
$\eps \ln \frac{\eps}{\delta} > 2/e.$ Define a probability measure $\nu$ on $L^2([0,1])$ by
$$ \nu(A) = \frac{|\mu|(\tilde{\lambda}^{-1}(A))}{\Tr{e^\Gamma}} \mbox{ for } 
A \in {\cal B}(L^2([0,1])). $$
There exists a compact set $K(\delta) \subset L^2([0,1])$ such that $\nu(K(\delta)^c) < \delta$.

Defining
\begin{equation} \label{Gepsdel} G(\eps,\delta) = \{ \alpha \in {\cal M}^+_1(L^2([0,1])):\, \alpha(K(\delta)^c) \leq \eps \}, \end{equation}
we claim that
\begin{equation} \label{Gbnd} \frac{1}{n} \ln \nu^{\otimes n} \left( \left\{
(\eta_k)_{k=1}^n:\, \frac{1}{n} \sum_{k=1}^n \delta_{\eta_k} \in G(\eps,\delta)^c \right\} \right) \leq
\frac{\eps}{2} \ln \frac{\delta}{\eps}. \end{equation}

To see this, note that
\begin{eqnarray*} \lefteqn{ \nu^{\otimes n} \left( \left\{
(\eta_k)_{k=1}^n:\, \frac{1}{n} \sum_{k=1}^n \delta_{\eta_i} \notin G(\eps,\delta) \right\} \right) =} \\
&=& \nu^{\otimes n} \left( \left\{ (\eta_k)_{k=1}^n:\, \frac{1}{n} \sum_{k=1}^n
\one_{K(\delta)^c}(\eta_k) > \eps \right\} \right)
\\ &=& (\beta_\delta)_n((\eps,+\infty)), \end{eqnarray*} where $\beta_\delta$ is the 
Bernoulli measure
$$ \beta_\delta = p_\delta \delta_1 + (1-p_\delta) \delta_0, \mbox{ with } 
p_\delta = \nu(K(\delta)^c) < \delta. $$
Here $(\beta_\delta)_n$ denotes the image measure $A_n(\beta_\delta^{\otimes n})$, 
where $A_n$ is the averaging map 
$$ A_n(x_1,\dots,x_n) = \frac{1}{n} \sum_{k=1}^n x_k. $$

By the Markov inequality,
\begin{eqnarray*} (\beta_\delta)_n((\eps,+\infty)) &\leq & 
e^{-n t \eps} \int_\RR e^{n t x} (\beta_\delta)_n(dx)  \\ &=& 
e^{-n t \eps} \int_{\RR^n} e^{n t \sum_{k=1}^n x_k} \prod_{k=1}^n \beta_\delta(dx_k)  \\
&=& e^{-n t \eps + n C_\delta(t)}, \end{eqnarray*}
where 
$$ C_\delta(t) = \ln \int e^{t x} \beta_\delta(dx) =  \ln(p_\delta e^t + 1 - p_\delta) $$ 
is the corresponding cumulant generating function. 
Taking $t  = \ln \frac{\eps(1-p_\delta)}{(1-\eps) p_\delta}$
(the maximiser of $t \eps- C_\delta(t)$), we find that 
\begin{equation} \label{GLDbnd} \nu^{\otimes n} \left( \left\{
(\xi_k)_{k=1}^n:\, \frac{1}{n} \sum_{k=1}^n \delta_{\xi_k} \notin G(\eps,\delta) \right\} \right) \leq 
e^{-n I_\delta(\eps)}, \end{equation} where
\begin{eqnarray} \label{Ideltabnd} 
I_\delta(\eps) &=& \sup_{t \in \RR}\, [t \eps - \ln(p_\delta e^t + (1-p_\delta))]  \non \\
&=& \eps \ln \frac{\eps}{p_\delta} + (1-\eps) \ln \frac{1-\eps}{1-p_\delta} \non \\
&>& \eps \ln \frac{\eps}{\delta} + (1-\eps) \ln (1-\eps) \non \\ &\geq & 
\eps \ln \frac{\eps}{\delta} + \frac{1}{e} > \half \eps \ln \frac{\eps}{\delta}.
\end{eqnarray}
Inserting this into (\ref{GLDbnd}) we obtain  (\ref{Gbnd}).

Now assume $\tilde{L} \geq 1$ and choose a sequence $\eps_l \downarrow 0$. 
Put $\delta_l = \eps_l \exp[-2 \tilde{L} l/\eps_l]$. 
Then ${\cal K}_L = \bigcap_{l=1}^\infty G_l \subset {\cal M}^+_1(L^2([0,1])$, with 
$G_l = G(\eps_l,\delta_l)$, is compact by Prokhorov's theorem, and
$$ \nu^{\otimes n} \left( \left\{ (\xi_k)_{k=1}^n:\,
\frac{1}{n} \sum_{k=1}^n \delta_{\xi_k} \in {\cal K}_{\tilde{L}}^c \right\}
\right) \leq \sum_{l=1}^\infty e^{-n \tilde{L}l} \leq 2 e^{-n \tilde{L}}. $$

Finally, consider the map $r: {\cal M}^+_1(L^2([0,1])) \to L^2([0,1])$ defined by 
$$ r(\nu) = \int_{L^2([0,1])} \psi\,\nu(d\psi), $$
where the integral is a Bochner integral. This integral is continuous on compacta, so the
set $\tilde{K}_{\tilde{L}} = r({\cal K}_{\tilde{L}})$ is also compact, and we have
\begin{eqnarray*} \lefteqn{\limsup_{n \to \infty} \frac{1}{n} \ln \nu^{\otimes n} 
\left( \left\{ (\eta_k)_{k=1}^n:\,
\frac{1}{n} \sum_{k=1}^n \xi_k \in \tilde{K}_{\tilde{L}}^c \right\} \right)} \\ &=& 
\limsup_{n \to \infty} \frac{1}{n} \nu^{\otimes n} \left( \left\{ (\eta_k)_{k=1}^n:\,
\frac{1}{n} \sum_{k=1}^n \delta_{\eta_k} \in {\cal K}_{\tilde{L}}^c \right\}
\right) < -\tilde{L}. \end{eqnarray*}
Replacing $\tilde{L}$ by $L = \tilde{L} + \Tr e^\Gamma$, we obtain the estimate (\ref{QLDtight}). \qed

Next, we need a generalization of the non-commutative H\"older inequality \cite{RS2}, 
Appendix to \S IX.4, Prop. 5. The generalized version is in \cite{Araki1973} but we give a proof in the appendix for completeness. (See also \cite{Dor2009}.)

\begin{lemma} \label{QLD-L3-3} Let $p_1,\dots,p_N \in (1,+\infty)$ ($N \in \NN$)
be such that $p_1^{-1} + \dots + p_N^{-1} = 1$.  If $A_k \in {\cal T}_{p_k}$ for $k=1,\dots,N$ 
then $\prod_{k=1}^N A_k \in {\cal T}_1$ and 
$$ \left\| \prod_{k=1}^N A_k \right\|_1 \leq \prod_{k=1}^N || A_k||_{p_k}. $$
\end{lemma}

We now write a Trotter product expansion as follows
\begin{eqnarray} \label{MFTrotter} \Tr e^{n[f(X^{(n)}) - H^{(n)}]} &=& 
\lim_{N \to \infty} \Tr \left[ \left(e^{n f(X^{(n)})/N} e^{-n H^{(n)}/N} \right)^N \right] \non \\
&=& \int e^{n \int_0^1 f(\frac{1}{n} \sum_{k=1}^n \xi_k(t))\,dt} \prod_{k=1}^n \kappa(d\xi_k).
\end{eqnarray} 
Given $\eps > 0$ we divide the interval $S$ into equal parts
$[a_i,a_{i+1}]$ ($i=1,\dots,r$) of length $|a_{i+1}-a_{i}| = \delta$ such that the variation of $f$ over each is less than $\eps$. Let $P_i$ be the projection onto the eigenspace of $X^{(n)}$ with 
eigenvalue in $[a_{i-1},a_{i}]$. We claim that the following LD upper bound holds.

\begin{lemma} \label{QLD-L3-4} Given $\eta > 0$, there exists $N_0 \in \NN$ such that for $N \geq N_0$
and $i_1,\dots, i_N \in \{1,\dots,r\}$ the following inequality holds.
\begin{eqnarray} \label{LDupper} \lefteqn{\frac{1}{n} \ln \left| \Tr 
\left[P_{i_N} e^{-n H^{(n)}/N} \dots P_{i_1} e^{-n H^{(n)}/N} \right] \right| } \non \\
&\leq & -\sum_{i=1}^r \gamma_i \inf_{a \in [a_{i-1},a_i]} I(a) + \ln \Tr e^{-H} + \eta, \end{eqnarray} 
where $\gamma_i$ is the fraction of $i_k = i$, i.e. $\gamma_i =  N_i/N$ if 
$N_i = \#\{k=1,\dots,N:\,i_k=i\}$.
\end{lemma}

\Pf To prove this, assume that the supremum in
\begin{equation} I(a) = \sup_{t \in \RR} [t a-C(t)] \end{equation}
is attained at $t=t(a)$. Assume that $I(c)=0$, i.e. $C'(t(c)) = c$. 
For $a_i < c$, $I(a)$ is decreasing for $a\leq a_i$ and hence $t(a) < 0$. 
Therefore $P_i \leq e^{n t(a_i)(X^{(n)}-a_i)/N}$. 
On the other hand, if $a_{i-1} > c$ then $t(a) > 0$ for $a > a_{i-1}$ and we have 
$P_i \leq e^{n t(a_{i-1})(X^{(n)}-a_{i-1})/N}$. 
If $c \in [a_{i-1},a_i]$ we simply write $P_i \leq \one$. 
(Note that in that case $t(c) = 0$.)
We set $b_i = a_{i-1}$ if $a_{i-1} > c$, $b_i = a_i$ if $a_i < c$ and $b_i = c$ if $a_{i-1} \leq c \leq a_i$.

Using Lemma~\ref{QLD-L3-3}, we have
\begin{equation} \left| \Tr \left[ P_{i_N} e^{-n H^{(n)}/N} \dots P_{i_1} 
e^{-n H^{(n)}/N} \right] \right| \leq 
\prod_{k=1}^N \left\{ \Tr \left[ \left( P_{i_k} e^{-n H^{(n)}/N} \right)^N \right] \right\}^{1/N}.
\end{equation}
Writing 
\begin{eqnarray*}  \Tr \left[ \left( P_{i_k} e^{-n H^{(n)}/N} \right)^N \right] &=&
\Tr \left[ \left( e^{-n H^{(n)}/2N} P_{i_k} e^{-n H^{(n)}/2N} \right)^N \right] \\
&\leq & \Tr \left[ \left( e^{-n H^{(n)}/2N} e^{n t(b_{i_k})(X^{(n)} - b_{i_k})/N} 
e^{-n H^{(n)}/2N} \right)^N \right], \end{eqnarray*} we obtain
\begin{eqnarray} \lefteqn{\left| \Tr \left[ P_{i_N} e^{-n H^{(n)}/N} \dots P_{i_1} 
e^{-n H^{(n)}/N} \right] \right|} \non \\ &\leq &
\prod_{k=1}^N \left\{ \Tr \left[ \left( e^{n t(b_{i_k})(X^{(n)} - b_{i_k})/N} e^{-n H^{(n)}/N} \right)^N \right] \right\}^{1/N} \non \\ &=& 
e^{-n \sum_{k=1}^N t(b_{i_k}) b_{i_k}/N} \prod_{k=1}^N \left\{ \Tr \left[ \left( e^{n t(b_{i_k}) X^{(n)}/N} e^{-n H^{(n)}/N}   \right)^N \right] \right\}^{1/N} \non \\ &=& 
e^{-n \sum_{k=1}^N t(b_{i_k}) b_{i_k}/N} \prod_{k=1}^N \prod_{j=1}^n 
\left\{ \Tr \left[ \left( e^{t(b_{i_k}) X_j/N} e^{-H_j/N}   \right)^N \right] \right\}^{1/N}  
\non \\ &=&  \prod_{i=1}^r e^{-n \gamma_i t(b_{i}) b_{i}}  \prod_{i=1}^r \prod_{j=1}^n 
\left\{ \Tr \left[ \left( e^{t(b_{i}) X/N} e^{-H/N}  \right)^N \right] \right\}^{\gamma_i}. 
\end{eqnarray}

Now, $-\lambda_- < b_1 < \dots < b_r < \lambda_+$, and therefore $\{t(b_i)\}_{i=1}^r$ is bounded.
By the proof of the Lie-Trotter theorem  there exists $N_0$ such that 
for $N \geq N_0$, 
$$ \Tr \left[ \left( e^{t(b_{i}) X/N} e^{-H/N}  \right)^N \right] \leq (1+\eta) \Tr [e^{t(b_i) X - H}]. $$
Inserting this, we have
\begin{eqnarray} \lefteqn{\left| \Tr \left[ P_{i_N} e^{-n H^{(n)}/N} \dots P_{i_1} 
e^{-n H^{(n)}/N} \right] \right|} \non \\ &\leq & 
\prod_{i=1}^r e^{-n \gamma_i t(b_{i}) b_{i}}  \prod_{i=1}^r \prod_{j=1}^n (1+\eta)^{\gamma_i}
\left\{ \Tr \left[ e^{t(b_{i}) X-H} \right] \right\}^{\gamma_i} \non \\ &=& 
 (1+\eta)^n \prod_{i=1}^r e^{-n \gamma_i t(b_{i}) b_{i}} 
e^{n \gamma_i C(t(b_i))} (\Tr\,e^{-H})^{n\gamma_i}
\end{eqnarray}
and therefore
\begin{eqnarray} \lefteqn{\frac{1}{n} \ln \left| \Tr \left[ P_{i_N} e^{-n H^{(n)}/N} \dots P_{i_1} 
e^{-n H^{(n)}/N} \right] \right|} \non \\
&\leq & \ln (1+\delta) - \sum_{i=1}^r \gamma_i (t(b_i) b_i - C(t(b_i)) - \ln \Tr e^{-H}) \non \\ &=& \ln(1+\eta)  - \sum_{i=1}^r \gamma_i I(b_i) + \ln \Tr e^{-H}. \end{eqnarray}
This proves the lemma. \qed

We are now ready to prove the upper bound for 
$\Tr \big(e^{n \big(f(X^{(n)}) - H^{(n)})} \big)$.

\begin{prop} \label{QLD-Prop1} The following large-deviation upper bound holds.
\begin{equation} \limsup_{n \to \infty} \frac{1}{n} \ln \Tr e^{n[f(X^{(n)}) - H^{(n)}]} 
\leq  \sup_{a \in [-||x||,||x||]} [f(a) - I(a)] + \ln \Tr(e^{-H}). \end{equation}
\end{prop}

\Pf First note that by Lemma~\ref{QLD-L3-1} there exists a compact set $K_L \subset L^2([0,1])$ such that 
\begin{equation} \limsup_{n \to \infty} \frac{1}{n} \ln |\mu|^{\otimes n} 
\left( \left\{ (\xi_k)_{k=1}^n:\,
\frac{1}{n} \sum_{k=1}^n \tilde{\lambda}(\xi_k) \in K_L^c \right\} \right) < -L. \end{equation}
Then 
\begin{eqnarray*} \lefteqn{\limsup_{n \to \infty} \frac{1}{n} \ln \Tr e^{n[f(X^{(n)}) - H^{(n)}]} } 
\non \\
&=& \limsup_{n \to \infty} \frac{1}{n} \ln \int_{L^2([0,1])^n} \exp \left\{ n \int_0^1 
f \big(\frac{1}{n} \sum_{k=1}^n \tilde{\lambda}(\xi_k(t)) \big) dt \right\} \non \\ && 
\qquad \qquad \times \exp \left\{- \sum_{k=1}^n \int_0^1 H_D(\xi_k(t)) \, dt \right\} 
\prod_{k=1}^n \mu(d\xi_k) \non \\ 
&=& \limsup_{n \to \infty} \frac{1}{n} \ln \bigg( \int_{A_n^{-1}(K_L)} \exp \left\{ n \int_0^1 
f \big(\frac{1}{n} \sum_{k=1}^n \tilde{\lambda}(\xi_k(t)) \big) dt \right\} \non \\ && 
\qquad \qquad \times \exp \left\{- \sum_{k=1}^n \int_0^1 H_D(\xi_k(t)) \, dt \right\} 
\prod_{k=1}^n \mu(d\xi_k) \non \\  && \qquad +
\int_{A_n^{-1}(K_L^c)} \exp \left\{ n \int_0^1 
f \big(\frac{1}{n} \sum_{k=1}^n \tilde{\lambda}(\xi_k(t)) \big) dt \right\} \non \\ && 
\qquad \qquad \times \exp \left\{- \sum_{k=1}^n \int_0^1 H_D(\xi_k(t)) \, dt \right\} 
\prod_{k=1}^n \mu(d\xi_k) \bigg). \end{eqnarray*} 

The second term is bounded by 
\begin{align*} &\limsup_{n \to \infty} \frac{1}{n} \ln \int_{A_n^{-1}(K_L^c)} \exp \left\{ n \int_0^1 
f \big(\frac{1}{n} \sum_{k=1}^n \tilde{\lambda}(\xi_k(t)) \big) dt \right\} \non \\ &\qquad \times 
\exp \left\{-\sum_{k=1}^n \int_0^1 H_D(\xi_k(t)) \, dt \right\}  \prod_{k=1}^n |\mu|(d\xi_k)
\\ &\qquad \leq  ||f||_\infty + ||H_D|| - L \end{align*}
and taking $L$ large enough, this is less than $\sup_{a \in S} [f(a) - I(a)] + \ln \Tr(e^{-H})$. 

It remains to show that 
\begin{align}  &\limsup_{n \to \infty} \frac{1}{n} \ln \left\{ \int_{K_L} 
\exp \left[ n \int_0^1 f (\eta(t)))\, dt \right] \kappa_n(d\eta) \right\} \non \\
&\qquad \leq  \sup_{a \in S} [f(a) - I(a)] + \ln \Tr(e^{-H}). \end{align}

For this, we introduce on $L^2([0,1])$ the Haar basis as in \cite{Dor2009}. 
Because $K_L$ is compact there exists for any $\eta > 0$, 
a finite $M \in \NN$ with $N = 2^M \geq N_0$ such that $||\eta - \pi_N(\eta)||_2 <\eta$, where 
$$ \pi_N(\eta) = \sum_{j=0}^{N-1} \langle h_j\,,\,\eta\rangle\, h_j. $$
Since the map $\xi \mapsto \int_0^1 f(\eta(t))\,dt$ is continuous $L^2([0,1]) \to \RR$, 
it suffices to prove that 
\begin{align}  &\limsup_{n \to \infty} \frac{1}{n} \ln \left\{ \int_{K_L} 
\exp \left[ n \int_0^1  f (\pi_N(\eta)(t))\, dt \right]  \kappa_n(d\eta) \right\}  \non \\
&\qquad \leq  \sup_{a \in S} [f(a) - I(a)] + \ln \Tr(e^{-H}). \end{align}
The path $\pi_N(\eta)$ is constant on intervals $[(k-1)/N, k/N]$. Therefore
$$ \int_0^1 f(\pi_N(\eta(t)) dt = \frac{1}{N} \sum_{k=1}^N f(\eta(k/N)) $$ and we can write
\begin{eqnarray} \lefteqn{\int_{K_L} 
\exp \left[ n \int_0^1  f (\pi_N(\eta)(t))\, dt \right]  \kappa_n(d\eta) } 
\non \\ &=& \sum_{j_1,\dots,j_N \in \NN_m^n} \prod_{k=1}^N e^{n f(\eta(j_k))/N}
\Tr \left[ P^1_{j_N} e^{- n H^{(n)}/N} \dots P^1_{j_1} e^{-n H^{(n)}/N} \right], \end{eqnarray}
where $P^1_{j_k}$ is the one-dimensional projection onto the eigenspace of $X^{(n)}$ with
eigenvalue $\frac{1}{n} \sum_{i=1}^n \lambda_{j_{k,i}}$ and
$$ \eta(j_k) = \frac{1}{n} \sum_{i=1}^n \tilde{\lambda}(j_{k,i}). $$
By continuity of $f$ this is bounded by
\begin{eqnarray} \lefteqn{\int_{K_L}  \exp \left[ n \int_0^1  f (\pi_N(\eta)(t))\, dt \right]  
\kappa_n(d\eta) } 
\non \\ &\leq& \sum_{i_1,\dots,i_N =1}^r \prod_{k=1}^N e^{n [f(b_{i_k}) + \eps]/N}
\left|\Tr \left[ P_{i_N} e^{- n H^{(n)}/N} \dots P_{i_1} e^{-n H^{(n)}/N} \right] \right|, 
\end{eqnarray}
Now applying Lemma~\ref{QLD-L3-4} we have
\begin{eqnarray} \lefteqn{\int_{K_L} 
\exp \left[ n \int_0^1  f (\pi_N(\eta)(t))\, dt \right]  \kappa_n(d\eta)} 
\non \\ &\leq& \sum_{\substack{N_1,\dots,N_r \geq 0: \\ \sum N_i = N}} \frac{N!}{N_1! \dots N_r!}\prod_{i=1}^r \bigg( e^{n N_i [f(b_{i}) + \eps]/N} e^{-n \gamma_i  
\sup_{a \in [a_{i-1},a_i]} I(a) +\eta} \bigg) (\Tr e^{-H})^n. 
\non \\ && \end{eqnarray}
Since $N$ is independent of $n$ we can take the limit $n \to \infty$ to get
\begin{eqnarray} \lefteqn{\limsup_{n \to \infty} \frac{1}{n} \ln \int_{K_L} 
\exp \left[ n \int_0^1  f (\pi_N(\eta)(t))\, dt \right]  \kappa_n(d\eta)} \non \\
&\leq & \sup_{a \in S} [f(a) - I(a)] + \ln \Tr(e^{-H}) + \eta + \eps. \end{eqnarray}
Since $\eps > 0 $ and $\eta > 0$ are arbitrary, we obtain the upper bound. \qed

\setcounter{equation}{0}

\section{The lower bound} 

To prove the reverse inequality, we first relate the rate function to the relative entropy.

Let $\cal S$ denote the set of states on $\cal M$, i.e. the linear maps $\varphi: {\cal M} \to \CC$ 
which are non-negative and unital: 
$$ A \geq 0 \implies \varphi(A) \geq 0; \quad \varphi(\one) = 1. $$
They are given by a density matrix $D_\varphi \in {\cal M}$ such that $\varphi(A) = \Tr(D_\varphi A)$.
Clearly, $D_\varphi \geq 0$ and $\Tr(D_\varphi) = 1$.
Given two states $\varphi, \rho \in {\cal S}$, the quantum relative entropy 
$S(\varphi\,||\,\rho)$ is defined by
\begin{equation} S(\varphi\,||\,\rho) = \Tr[D_\varphi\,\ln(D_\varphi)] - \Tr[D_\varphi\,\ln(D_\rho)], \end{equation}
where $D_\varphi$ and $D_\rho$ are the density matrices for $\varphi$ and $\rho$ respectively.

It is well-known (see for example \cite{Lieb1973},  \cite{NC2000} or \cite{Israel1979}) that 
$S(\varphi\,||\,\rho)$ has the following properties. 

\begin{lemma} \label{QLD-L4-1}  The relative entropy $S(\varphi\,||\,\rho)$ is non-negative and convex 
jointly in in $\varphi$ and $\rho$. \end{lemma}

Next we need some standard results about the free energy.

\begin{lemma} \label{QLD-L4-2} Let $X,H \in {\cal M}$ be hermitian. Then 
$$ \ln \frac{\Tr[e^{tX - H}]}{\Tr[e^{-H}]} \geq t \varphi(X) - S(\varphi\,||\,\rho) $$
for every state $\varphi \in {\cal S}$, where $\rho$ is the Gibbs state
$\rho(A) = \Tr[A\,e^{-H}]/\Tr[e^{-H}]$. Moreover, equality holds only when $\varphi$ is 
the perturbed Gibbs state $\omega_t$ given by
$$ \omega_t(A) = \frac{\Tr[A\,e^{tX-H}]}{\Tr[e^{tX-H}]}. $$
\end{lemma}

%\Pf We have, noting that 
%$$ \rho = \frac{e^{-H}}{\Tr(e^{-H})} \mbox{ and } 
%D_{\omega_t} = \frac{e^{tX -H}}{\Tr(e^{tX-H})}, $$
%\begin{eqnarray*} t\varphi(X) - S(\varphi\,||\,\rho) &=& \Tr \left[ D_\varphi (t X - \ln D_\varphi + \ln %D_\rho) \right] \\ &=& \Tr \left[ D_\varphi (tX - H - \ln D_\varphi) \right] - \ln \Tr (e^{-H}) \\
%&=& \Tr \left[ D_\varphi (\ln D_{\omega_t} - \ln D_\varphi) \right] + \ln \Tr e^{tX-H}
%- \ln \Tr e^{-H} \\ &=& \ln \frac{\Tr e^{tX-H}}{\Tr e^{-H}} - S(\varphi\,||\,\omega_t) 
%\leq  \ln \frac{\Tr e^{tX-H}}{\Tr e^{-H}} . \end{eqnarray*} 
%This proves the inequality. Moreover, since equality in Lemma~\ref{QLD-L5} only holds if 
%$A=B$ if $F$ is strictly convex, it follows that 
%$S(\varphi\,||\,\omega_t) > 0$ unless $\varphi = \omega_t$.  \qed

We now repeat two lemmas from \cite{PRV1989}. 
\begin{lemma} \label{QLD-L4-3} Let $\rho$ be the Gibbs state with density 
$D_\rho = e^{-H}/\Tr[e^{-H}]$, with $H=H^* \in {\cal M}$.
For every state $\varphi \in {\cal S}$ on $\cal M$, 
$$ S(\varphi\,||\,\rho) \geq I(\varphi(X)) \mbox{ for all } X=X^* \in {\cal M}. $$ 
Moreover, if $S(\varphi\,||\,\rho) = I(\varphi(X)) < +\infty$ then either there is a $t \in \RR$ such that
$$ \varphi(A) = \omega_t(A) = \frac{\Tr(A\, e^{tX-H})}{\Tr(e^{tX-H})}, $$
or else $D_\varphi$ is the projection onto $\lambda_+$ or $\lambda_-$.
\end{lemma}

\Pf  By Lemma~\ref{QLD-L4-2}, we have, for any hermitian $X \in {\cal M}$ and $t \in \RR$,
\begin{eqnarray*} S(\varphi\,||\,\rho) &\geq &  
t\varphi(X) - \ln \Tr(e^{tX-H}) + \ln \Tr(e^{-H}) \\&=&  
t \varphi(X) - C(t).  \end{eqnarray*}
Therefore, $$ S(\varphi\,||\,\rho) \geq \sup_{t \in \RR} [t \varphi(X) - C(t)] = I(\varphi(X)). $$
If $\varphi = \omega_t$ then $S(\varphi\,||\,\rho) = t \varphi(X) - \ln \Tr e^{tX-H} + \ln \Tr e^{-H} = 
t \varphi(X) - C(t) \leq I(\varphi(X))$ so that equality holds. Conversely, if $S(\varphi\,||\,\rho) = I(\varphi(X)) < +\infty$ then suppose first that $I(\varphi(X)) = t_0 \varphi(X) - C(t_0)$ 
where $t_0$ is the unique solution of $\varphi(X) = C'(t)$. 
Then by the uniqueness in Lemma~\ref{QLD-L4-2},
$$ S(\varphi\,||\,\rho) = I(\varphi(X)) = t_0\varphi(X) - \ln \Tr(e^{t_0 X-H}) + \ln \Tr(e^{-H})  $$
implies that $\varphi = \omega_{t_0}$. Otherwise, $I(\varphi(X)) = \lim_{t \to \pm \infty} [t \varphi(X) - C(t)]$. For large $|t|$, $C(t) \sim t \lambda_\pm - \Tr(P_\pm H) - \ln \Tr e^{-H}$, where $P_\pm$ 
are the projections onto the eigenstates of $X$ corresponding to $\lambda_\pm$. Hence
$$ I(\varphi(X)) =  \Tr(P_\pm H) + \ln \Tr e^{-H} = S(P_\pm\,||\,\rho). $$
 \qed

\begin{cor} \label{QLD-cor2}
For any continuous function $f:[-||X||,||X||] \to \RR$, the following identity holds.
$$ \sup_{u \in {\rm co}(\sigma(X))} \{f(u) - I(u)\} = 
\sup_{\varphi \in {\cal S}} \{f(\varphi(X)) - S(\varphi\,||\,\rho) \}. $$
\end{cor}

\Pf By Lemma~\ref{QLD-L4-3}, 
\begin{eqnarray*} \sup_{\varphi \in {\cal S}} \{f(\varphi(X)) - S(\varphi\,||\,\rho) \} &\leq & 
\sup_{\varphi \in {\cal S}} \{ f(\varphi(X)) - I(\varphi(X))\} \\ 
&\leq & \sup_{u \in [-||X||,||X||]} \{f(u) - I(u)\}. \end{eqnarray*}
To prove the reverse inequality, we may assume that $I(u) < +\infty$.
On the other hand, let $t(u)$ be such that $u = C'(t(u))$, and put $\varphi = \omega_{t(u)}$. 
Then $S(\varphi\,||\,\rho) = t(u) \varphi(X) - C(t(u)) = I(u)$ and hence
$$ \sup_{\varphi \in {\cal S}} \{f(\varphi(X)) - S(\varphi\,||\,\rho) \} \geq f(u) - I(u) $$ 
and since this holds for all $u \in {\rm co}(\sigma(X))$ this implies the reverse inequality. \qed

To prove the lower bound, we need one more standard inequality.

\begin{lemma} \label{QLD-L4-4} If $A$ and $B$ are hermitian matrices then 
$$ \left| \ln \Tr(e^A) - \ln \Tr(e^B) \right| \leq ||A-B||. $$
\end{lemma} 

%\Pf Let $A(t) = B + t(A-B)$. Then 
%\begin{eqnarray*} \ln \Tr e^A - \ln \Tr e^B &=& \int_0^1 dt \frac{d}{dt} \ln \Tr e^{A(t)} \\
%&=& \int_0^1 dt \frac{\Tr[(A-B) e^{A(t)}]}{\Tr[e^{A(t)}]} \end{eqnarray*}
%and the inequality follows from the general fact that $\Tr(AB) \leq ||A||\,\Tr(B)$. \qed

We are now ready to prove the lower bound.
\begin{prop} \label{QLD-Prop2}
If $X, H \in {\cal M}$ are hermitian matrices and \\ $f: [-||X||,||X||] \to \RR$  is continuous then 
$$ \liminf_{n \to \infty} \frac{1}{n} \ln \Tr e^{n(f(X^{(n)}) - H^{(n)})} 
\geq \sup_{a \in [-||X||,||X||]} \{f(a) - I(a)\} + \ln \Tr e^{-H}. $$
\end{prop} 

\Pf First note that by Lemma~\ref{QLD-L4-4}, we can assume that $f$ is a polynomial. Indeed, 
if $\eps > 0$ then there is a polynomial $P$ such that $\sup_{a \in [-||X||,||X||]} |f(a) - P(a)| < \eps$. 
Then 
$$ \frac{1}{n} \left| \ln \Tr e^{n(f(X^{(n)}) - H^{(n)})} - 
\ln \Tr e^{n(P(X^{(n)}) - H^{(n)})} \right| \leq \eps $$ and 
$$ \left| \sup_{a \in [-||X||,||X||]} \{f(a) - I(a)\} - \sup_{a \in [-||X||,||X||]} \{P(a) - I(a)\} \right| \leq \eps. $$
Similarly, consider a monomial $P_k(x) = x^k$. Then 
\begin{eqnarray*} \lefteqn{\| P_k(X^{(n)}) - \frac{k!}{n^k}  \sum_{i_1 < \dots < i_k} X_{i_1} \dots X_{i_k} \|} \\  &\leq &
\frac{n^k - k! {n \choose k}}{n^k} ||X||^k = 
\left(1 - (1-\frac{1}{n}) \dots (1-\frac{n-k+1}{n}) \right) ||X||^k. \end{eqnarray*}
Since $(1-\frac{1}{n}) \dots (1-\frac{n-k+1}{n}) \to 1$, it follows that we can replace the polynomial 
$P(X^{(n)}) = \sum_{k=0}^r c_k (X^{(n)})^k$ by 
\begin{equation} \label{gapprox} g(X_1,\dots,X_n) = \sum_{k=0}^r c_k \frac{k!}{n^k}  
\sum_{i_1 < \dots < i_k} X_{i_1} \dots X_{i_k}. \end{equation} 
Now using Lemma~\ref{QLD-L4-2}, we have 
\begin{eqnarray} \lefteqn{\frac{1}{n} \ln \Tr e^{n(g(X_1,\dots,X_n) - H^{(n)})}} \non \\ &\geq  &
\frac{1}{n} \sup_{\varphi \in {\cal S}_n} 
\left\{n \varphi(g(X_1,\dots,X_n))  - S(\varphi\,||\,\rho^{\otimes n})\right\} + 
\frac{1}{n} \ln \Tr e^{-n H^{(n)}}. \end{eqnarray}
(Here ${\cal S}_n$ is the state space of ${\cal M}^{\otimes n}$.)
In the supremum, we can restrict the states to product states, $\varphi = \omega^{\otimes n}$. 
Hence
\begin{eqnarray} \lefteqn{\frac{1}{n} \ln \Tr e^{n(g(X_1,\dots,X_n) - H^{(n)})}} \non \\ &\geq  &
\sup_{\omega \in {\cal S}} 
 \left\{ \omega^{\otimes n}\left( \sum_{k=0}^r c_k \frac{k!}{n^k}  
\sum_{i_1 < \dots < i_k} X_{i_1} \dots X_{i_k} \right)  \right. \non \\
&& \qquad \left. - \frac{1}{n}  S(\omega^{\otimes n}\,||\,\rho^{\otimes n})\right\} +  \ln \Tr e^{-H}. 
\non \\  &=& \sup_{\omega \in {\cal S}} \left\{  \sum_{k=0}^r c_k \frac{k!}{n^k}  
\sum_{i_1 < \dots < i_k} \omega(X)^k - S(\omega\,||\,\rho)\right\} +  \ln \Tr e^{-H}. 
\end{eqnarray}
Taking the limit, we have, using again that ${n \choose k} \sim n^k/k!$, 
\begin{eqnarray} \lefteqn{\liminf_{n \to \infty} 
\frac{1}{n} \ln \Tr e^{n(g(X_1,\dots,X_n) - H^{(n)})}} \non \\ &\geq  &
\sup_{\omega \in {\cal S}}  \left\{  \sum_{k=0}^r c_k \omega(X)^k - S(\omega\,||\,\rho)\right\} 
+  \ln \Tr e^{-H} \non \\ 
&=& \sup_{\omega \in {\cal S}} \{P(\omega(X)) - S(\omega\,||\,\rho)\} + \ln \Tr e^{-H}.
\non \\ &=& \sup_{u \in \ol{\rm co}(\sigma(X))} [P(u) - I(u)] + \ln \Tr(e^{-H}) \end{eqnarray}
by Corollary~\ref{QLD-cor2}.
This proves the lower bound for polynomials $P$ and hence for general continuous functions $f$. \qed
\medskip

\textbf{Example.} A typical example to which the PRV theorem applies is the mean-field transverse-field Ising model, with Hamiltonian given by 
\begin{equation} \label{TFIsing}  H_n = -\frac{1}{n} \sum_{i,j=1}^n \sigma^z_i \sigma_j^z - 
h \sum_{i=1}^n \sigma_i^x, \end{equation}
where $\sigma^z_i$ and $\sigma^x_i$ are Pauli matrices at position $i$.
The free energy density at inverse temperature $\beta > 0$ is given by 
\begin{equation} \label{Isingfe} 
f(\beta,h) = -\frac{1}{\beta} \lim_{n \to \infty} \frac{1}{n} \ln \Tr e^{-\beta H_n}. 
\end{equation}
The corresponding cumulant generating function is 
\begin{eqnarray*} C(t) &=& \ln \Tr e^{t \sigma^z + \beta h \sigma^x} - \ln \Tr e^{\beta h \sigma^x} 
\\ &=& \ln \cosh \sqrt{t^2 + \beta ^2 h^2} - \ln \cosh (\beta h). \end{eqnarray*}
Therefore, 
\begin{equation} \label{TFIsingrate} 
I(z) = \sup_{t \in \RR} \big( tz - \ln \cosh \sqrt{t^2+\beta^2 h^2}\big) + \ln \cosh (\beta h), \end{equation}
and 
\begin{equation} \label{TFIsingvar} f(\beta,h) = \inf_{z \in [-1,1]} \big[-z^2 + 
\frac{1}{\beta} \tilde{I}(z)],
\end{equation}
where 
\begin{equation} \tilde{I}(z) = I(z) - \ln 2\cosh (\beta h) = \sup_{t \in \RR} 
\big( tz - \ln 2\cosh \sqrt{t^2 + \beta^2 h^2}]. \end{equation}
(Note that $|C'(t)| \leq 1$.)

\setcounter{equation}{0}

\section{Two-variable generalization}

Lemma~\ref{QLD-L3-4} suggests that we can generalize Theorem~\ref{PRVThm} by replacing 
$e^{-n H^{(n)}/N}$ by projections $Q_j$ corresponding to the operator $H^{(n)}$.
We should then be able to consider functions of $H^{(n)}$ as well as $X^{(n)}$ which puts the 
two operators on an equal footing. In the following we write $Y$ instead of $H$.

We need the analogue of Lemma~\ref{QLD-L3-4}. The cumulant generating function is
\begin{equation} C(t_1,t_2) = \ln \Tr e^{t_1 X + t_2 Y}. \end{equation}
(Note that this is not normalized, i.e. $C(0,0)=\ln m \neq 0$.)
Then 
\begin{equation} I(x_1,x_2) = \sup_{t_1,t_2} [t_1 x_1 + t_2 x_2 - C(t_1,t_2)]. \end{equation}
We subdivide $S_1 = {\rm co}(\sigma(X))$ and $S_2 = {\rm co}(\sigma(Y))$ into small intervals 
of size $\delta$ such that the variation of $f$ and $g$ on these intervals is less than $\eps > 0$.

\begin{lemma} \label{QLD-L5-1} Given $\eta > 0$, there exists $N_0 \in \NN$ independent of $n$
such that for $N \geq N_0$ and $i_1,\dots,i_N \in \{1,\dots,r_1\}$ and $j_1,\dots,j_N \in \{1,\dots, r_2\}$, where $r_1 = |S_1|/\delta$ and $r_2 = |S_2|/\delta$, the following holds.
\begin{equation} \frac{1}{n} \ln \left| \Tr \left[ P_{i_N} Q_{j_N} \dots P_{i_1} Q_{j_1} \right] \right|
\leq - \sum_{i=1}^{r_1} \sum_{j=1}^{r_2} \gamma_{i,j} \inf_{\substack{x \in [x_{i-1},x_i] \\
y \in [y_{i-1},y_i]}} I(x,y) + \eta, \end{equation} 
where $\gamma_{i,j}$ is the fraction of $k \in \{1,\dots,N\}$ such that $i_k=i$ and $j_k=j$.
\end{lemma}

\Pf Assume that the maximum in $I(x,y) = \sup_{t_1,t_2} [t_1 x + t_2 y - C(t_1,t_2)]$
is attained at $(t_1(x,y), t_2(x,y))$. If $x \mapsto I(x,y)$ is minimal at $x=c_1(y)$ then 
$I(x,y)$ is decreasing for $x_i < c_1(y)$, and increasing for $x > c_1(y)$. Similarly, 
$y \mapsto I(x,y)$ is decreasing for $y < c_2(x)$ and increasing for $y > c_2(x)$. If $x_i < c_1(y_j)$ 
we set $a_{i,j} = x_i$ and if $x_{i-1} > c_1(y_j)$ we put $a_{i,j} = x_{i-1}$. 
If $c_1(y_j) \in [x_{i-1},x_i]$ we set $a_{i,j} = c(y_j)$. Similarly, if $y_j < c_2(x_i)$ then we set 
$b_{i,j} = y_j$ and if $y_{j-1} > c_2(x_i)$ we set $b_{i,j} = y_{j-1}$. 
Finally, if $c_2(x_i) \in [y_{j-1},y_j]$ then we set $b_{i,j} = c_2(x_i)$. 

Then we have
$$ P_{i_k} \leq e^{n t_1(a_{i_k,j_k},b_{i_k,j_k}) (X^{(n)} - a_{i_k,j_k})/N} \mbox{ and } 
Q_{j_k} \leq  e^{n t_2(a_{i_k,j_k},b_{i_k,j_k}) (Y^{(n)} - b_{i_k,j_k})/N}. $$ 
By Lemma~\ref{QLD-L3-4}, 
\begin{equation}  \left| \Tr \left[ P_{i_N} Q_{j_N} \dots P_{i_1} Q_{j_1} \right] \right| \leq 
\prod_{k=1}^N \left\{ \Tr \left[ \big( P_{i_k} Q_{j_k} \big)^N \right] \right\}^{1/N}, \end{equation}
where 
\begin{eqnarray} \Tr \left[ \big( P_{i_k} Q_{j_k} \big)^N \right] &=&
\Tr \left[ \big( P_{i_k} Q_{j_k} P_{i_k} \big)^N \right]  \non \\ &\leq & 
\Tr \left[ \big( P_{i_k}e^{n t_2(a_{i_k,j_k},b_{i_k,j_k}) (Y^{(n)} - b_{i_k,j_k})/N} 
P_{i_k} \big)^N \right] \non \\ &\leq & 
\Tr \bigg[ \big(e^{n t_1(a_{i_k,j_k},b_{i_k,j_k}) (X^{(n)} - a_{i_k,j_k})/N} \non \\  && \qquad \times
e^{n t_2(a_{i_k,j_k},b_{i_k,j_k}) (Y^{(n)} - b_{i_k,j_k})/N} \big)^N \bigg]. \end{eqnarray} 
Therefore
\begin{eqnarray} \lefteqn{\left|\Tr \left[ P_{i_N} Q_{j_N} \dots P_{i_1} Q_{j_1} \right] \right|} 
\non \\ &\leq & \prod_{k=1}^N \left\{ \Tr \left[ \big(
e^{n t_1(a_{i_k,j_k},b_{i_k,j_k}) (X^{(n)} - a_{i_k,j_k})/N} \right. \right. \non \\ 
&&  \qquad \times \left. \left.
e^{n t_2(a_{i_k,j_k},b_{i_k,j_k}) (Y^{(n)} - b_{i_k,j_k})/N} \big)^N \right] \right\}^{1/N} \non \\ 
&=& \prod_{k=1}^N  e^{-n \big( t_1(a_{i_k,j_k},b_{i_k,j_k}) a_{i_k,j_k} + 
t_2(a_{i_k,j_k},b_{i_k,j_k}) b_{i_k,j_k} \big)/N} 
\non \\ &&  \times \prod_{k=1}^N  \left\{\Tr \left[ 
\big(e^{n t_1(a_{i_k,j_k},b_{i_k,j_k}) X^{(n)}/N}  
e^{n t_2(a_{i_k,j_k},b_{i_k,j_k}) Y^{(n)}/N} \big)^N \right] \right\}^{1/N} 
\non \\ &=& \prod_{i=1}^{r_1} \prod_{j=1}^{r_2} 
e^{-n \gamma_{i,j} (t_1(a_{i,j},b_{i,j}) a_{i,j} + t_2(a_{i,j},b_{i,j}) b_{i,j})} 
\non \\ && \times  \prod_{i=1}^{r_1} \prod_{j=1}^{r_2} 
\left\{ \Tr \left[ \big(e^{t_1(a_{i,j},b_{i,j}) X/N}  
e^{t_2(a_{i,j},b_{i,j}) Y/N} \big)^N \right] \right\}^{n \gamma_{i,j}}. \end{eqnarray}
Now, by the Lie-Trotter theorem, given $\eta > 0$, there exists $N_0 \in \NN$ (independent of $n$) 
such that for $N \geq N_0$,
\begin{eqnarray*} \lefteqn{\Tr \left[ \big(e^{t_1(a_{i,j},b_{i,j}) X/N}  
e^{t_2(a_{i,j},b_{i,j}) Y/N} \big)^N \right]} \\  &\leq & (1+\eta) 
\Tr \left[ e^{t_1(a_{i,j},b_{i,j}) X + t_2(a_{i,j},b_{i,j}) Y} \right]. \end{eqnarray*}  
Therefore
\begin{eqnarray} \lefteqn{\left| \Tr \left[ P_{i_N} Q_{j_N} \dots P_{i_1} Q_{j_1} \right] \right|} 
\non \\ &\leq & \prod_{i=1}^{r_1} \prod_{j=1}^{r_2}  
e^{-n \gamma_{i,j} (t_1(a_{i,j},b_{i,j}) a_{i,j} + t_2(a_{i,j},b_{i,j}) b_{i,j})}
(1+\eta)^{n \gamma_{i,j}} \non \\ &&\qquad \times \prod_{i=1}^{r_1} \prod_{j=1}^{r_2} 
\left\{\Tr \left[ e^{t_1(a_{i,j},b_{i,j}) X + t_2(a_{i,j},b_{i,j}) Y} \right] \right\}^{n \gamma_{i,j}} 
\non \\ &=& (1+\eta)^n \prod_{i=1}^{r_1} \prod_{j=1}^{r_2}  
\left\{e^{-n \gamma_{i,j} (t_1(a_{i,j},b_{i,j}) a_{i,j} + t_2(a_{i,j},b_{i,j}) b_{i,j})}
\right. \non \\ && \qquad \times \left. e^{n \gamma_{i,j} C(t_1(a_{i,j},b_{i,j}), t_2(a_{i,j},b_{i,j}))} \right\} \non \\ &=&  (1+\eta)^n \prod_{i=1}^{r_1} \prod_{j=1}^{r_2}  
e^{-n \gamma_{i,j} I(a_{i,j},b_{i,j})}. \end{eqnarray}
Taking logarithms and dividing by $n$, the result follows. \qed

In order to interchange the limits $N \to \infty$ and $n \to \infty$ we need to introduce another QSP. 
Let $H_n \in {\cal M}^{\otimes n}$ be a general hermitian matrix with matrix elements 
$(H_n)_{\ul{k}, \ul{k}'}$, where $\ul{k}, \ul{k}' \in \NN_m^n$. As in Section~\S 2, we introduce 
\begin{equation} H_D(\ul{k}) = (H_n)_{\ul{k},\ul{k}} - 
\sum_{\ul{k}' \neq \ul{k}} |(H_n)_{\ul{k},\ul{k}'}|, \end{equation} and 
\begin{equation} (\tilde{H}_n)_{\ul{k},\ul{k}'} = (H_n)_{\ul{k},\ul{k}'} - H_D(\ul{k}) \delta_{\ul{k},\ul{k}'}. \end{equation} 
Given a subdivision $\sigma: 0 \leq t_1 < \dots < t_N \leq 1$, we define a complex-valued measure on $(\NN_m^n)^\sigma$ by 
\begin{eqnarray} \mu_n^\sigma(A_1 \times \dots \times A_N) &=& \sum_{\ul{k}_1 \in A_1} \dots 
\sum_{\ul{k}_N \in A_N} \left( e^{-(1-t_N+t_1) \tilde{H}_n} \right)_{\ul{k}_1, \ul{k}_N} \non \\ 
&& \times \left( e^{-(t_N-t_{N-1}) \tilde{H}_n} \right)_{\ul{k}_N, \ul{k}_{N-1}} \dots 
\left( e^{-(t_2-t_1) \tilde{H}_n} \right)_{\ul{k}_2, \ul{k}_1} \end{eqnarray}
for subsets $A_1, \dots, A_N \subset \NN_m^n$. As in Theorem~\ref{QSPThm}, these measures form 
a projective system, and the projective limit is a complex-valued measure $\mu_n$ on 
$(\NN_m^n)^{[0,1]}$. Moreover, $|\mu_n|$ has a generating matrix $e^{t \Gamma_n}$, where $\Gamma_n$ is given by 
\begin{equation} (\Gamma_n)_{\ul{k}, \ul{k}'} = \begin{cases} -|(\tilde{H}_n)_{\ul{k},\ul{k}}| =
H_D(\ul{k}) - (H_n)_{\ul{k},\ul{k}} &\text{if $\ul{k}' = \ul{k}$;} \\ |(H_n)_{\ul{k},\ul{k}'}| 
&\text{if $\ul{k}' \neq \ul{k}$.} \end{cases} \end{equation}
We now need a strengthened version of the concentration lemma, Lemma~\ref{QLD-L2-1}.

\begin{lemma} \label{QLD-L5-2} Consider the submatrix $\Gamma^{(i)}_n$ of $\Gamma_n$ for $i = 1,\dots,n$ defined by \begin{equation} (\Gamma^{(i)}_n)_{\ul{k},\ul{k}'} = 
\begin{cases} (\Gamma_n)_{\ul{k},\ul{k}'} &\text{if $k'_i \neq k_i$;} \\ 0 &\text{otherwise.} 
\end{cases} \end{equation}
Assume that there exists a constant $C$ independent of $n$ such that $|| \Gamma^{(i)}_n|| \leq C$
for all $i=1,\dots,n$. Define the probability measure $\nu_n$ by 
$$ \nu_n(A) = \frac{|\mu_n|(A)}{\Tr(e^{\Gamma_n})}. $$
Then for all $\delta > 0$ there exists a compact set $K(\delta) \subset D([0,1],\NN_m)$ independent of $n$ such that 
\begin{equation} \nu_n (\{(\xi_1,\dots,\xi_n) \in D([0,1],\NN_m^n):\, \xi_i \in K(\delta)^c\}) < \delta. \end{equation}
\end{lemma}

\Pf As in the proof of Lemma~\ref{QLD-L2-1}, we estimate the probability that $\xi_i$ makes at least two jumps in a small interval $[t_{j_1}, t_{j_2}]$. Analogous to equation (\ref{2jumpprob}) we have
\begin{eqnarray} \lefteqn{\pi_\sigma(|\mu_n|)(\{\xi_i \mbox{ makes at least 2 jumps in } [t_{j_1},t_{j_2}]\})} \non \\ &\leq & \sum_{j=j_1}^{j_2-1} \sum_{j'=j+1}^{j_2} \sum_{\ul{k}, \ul{k}', \ul{k}''} 
\sum_{\ul{l}:\, l_i \neq k'_i} \sum_{\ul{l}':\, l'_i \neq k''_i} (e^{(1-t_{j'}) \Gamma_n})_{\ul{k}, \ul{l}'} 
(e^{(t_{j'}-t_{j'-1}) \Gamma_n})_{\ul{l}', \ul{k}''} \non \\ && \times 
(e^{(t_{j'-1}-t_j) \Gamma_n})_{\ul{k}'', \ul{l}} (e^{(t_j-t_{j-1}) \Gamma_n})_{\ul{l}, \ul{k}'}
(e^{t_{j-1} \Gamma_n})_{\ul{k}', \ul{k}}. \end{eqnarray}
Now, for small $\delta t$, we have that if $l_i \neq k'_i$ then
\begin{equation} (e^{\delta t\,\Gamma_n})_{\ul{l},\ul{k}'} \sim \delta t \,(\Gamma^{(i)}_n)_{\ul{l},\ul{k}'} + O(\delta t^2). \end{equation} 
Therefore 
\begin{eqnarray} \lefteqn{\pi_\sigma(|\mu_n|)(\{\xi_i \mbox{ makes at least 2 jumps in } [t_{j_1},t_{j_2}]\})} \non \\ &\leq & \sum_{j=j_1}^{j_2-1} \sum_{j'=j+1}^{j_2} 
(t_j-t_{j-1}) (t_{j'}-t_{j'-1}) \non \\ && \qquad \qquad \times 
\Tr \left[ (e^{(1-t_{j'}) \Gamma_n})\, \Gamma^{(i)}_n\,
(e^{(t_{j'-1}-t_j) \Gamma_n})\, \Gamma^{(i)}_n\, (e^{t_{j-1} \Gamma_n}) \right]
\non \\ &\leq & C^2 \Tr(e^{\Gamma_n}) \delta^2, \end{eqnarray}
where $\delta = t_{j_2} - t_{j_1}$. 
Defining, as in the proof of Lemma~\ref{QLD-L2-1}, 
\begin{equation} G_\delta = \{ \xi \in D([0,1],\NN_m);\, \tilde{\omega}_{\delta}(\xi) \leq \eta\}) \end{equation}
and taking $\eta < 1$, we have, summing over the intervals $[t_{j_1}, t_{j_2}]$, 
\begin{equation} \pi_\sigma(\nu_n)(\pi_\sigma^{-1}(\{\xi: \,\xi_i \in G_\delta^c\})) \leq C^2 \delta, 
\end{equation} provided that the mesh of $\sigma$ is fine enough, i.e. $\max_{j=1}^N \{t_j-t_{j-1}\}$
is small enough. 
Then writing $K(\delta) = \bigcap_{k \in \NN} G_{\delta/k^2}$, we have that
\begin{equation} \pi_\sigma(\nu_n) (\pi_\sigma^{-1}(\{\xi:\, \xi_i \in K(\delta)^c\})) 
\leq \frac{\pi^2}{6} C^2 \delta. \end{equation} 
Finally, replace $\delta$ by $6 \delta/(\pi^2 C^2)$. \qed

We need a slight improvement on this lemma. Namely, in this general case, $\xi_i$ and $\xi_j$ are not independent. However, the probability that they jump at the same time is small. 
Therefore, the analogue of Lemma~\ref{QLD-L3-1} nevertheless holds. 

\begin{lemma} \label{QLD-L5-3} Consider the submatrices $\Gamma^{(I)}_n$ of $\Gamma_n$ for any finite $I \subset \{1,\dots,n\}$ defined by 
\begin{equation} (\Gamma^{(I)}_n)_{\ul{k},\ul{k}'} = 
\begin{cases} (\Gamma_n)_{\ul{k},\ul{k}'} &\text{if $k'_i \neq k_i\,$ for all $i \in I$;} \\ 0 &\text{otherwise.} 
\end{cases} \end{equation}
Assume that there exists a constant $C$ independent of $n$ such that 
$|| \Gamma^{(I)}_n|| \leq C^{|I|}$ for all $I \subset \{1,\dots,n\}$. 
Then for all $\delta > 0$ there exists a compact set $K(\delta) \subset D([0,1],\NN_m)$ independent of $n$ such that 
\begin{equation} \nu_n (\{(\xi_1,\dots,\xi_n) \in D([0,1],\NN_m^n):\, (\forall i \in I) \xi_i \in K(\delta)^c\}) < \delta^{|I|}. \end{equation}
\end{lemma}

\Pf This is proved in the same way as the previous lemma. For example, for $p=2$, in the expression for
$$ \pi_\sigma(|\mu_n|)(\{\xi_i \mbox{ and } \xi_j \mbox{ both make at least 2 jumps in } [t_{j_1},t_{j_2}]\}) $$ all jumps occur at different points, in which case there appear two factors 
$\Gamma^{(i)}_n$ and two factors $\Gamma^{(j)}_n$ each, or there is one pair of jumps and two 
separate jumps, resulting in a factor $\Gamma^{(i,j)}_n$ as well as one factor $\Gamma^{(i)}_n$ and 
one factor $\Gamma^{(j)}_n$ each, or there are two pairs of jumps, in which case $\Gamma^{(i,j)}_n$ 
occurs twice. By assumption, however, $ ||\Gamma^{(i,j)}_n || \leq C^2$ so we get
\begin{equation*}\pi_\sigma(|\mu_n|)(\{\xi_i \mbox{ and } \xi_j \mbox{ both make at least 2 jumps in } [t_{j_1},t_{j_2}]\}) \leq  C^4 \delta^4.  \end{equation*}
For disjoint intervals $[t_{j_1},t_{j_2}]$ and $[t_{j'_1},t_{j'_2}]$ it follows from Lemma~\ref{QLD-L5-2}
that \begin{align*}&\pi_\sigma(|\mu_n|)(\{\xi_i \mbox{ and } \xi_j \mbox{ make at least} \\  
&\qquad \mbox{2 jumps in } [t_{j_1},t_{j_2}] \mbox{ and } [t_{j'_1},t_{j'_2}] \mbox{ resp.}\}) \leq  C^4 \delta^4.  \end{align*}
More generally, for a finite set $I \subset \{1,\dots,n\}$,
\begin{equation}\pi_\sigma(|\mu_n|) \left(\bigcap_{i \in I} \{\xi_i \mbox{ makes at least 2 jumps in } [t_{j_i},t_{j'_i}]\} \right) \leq  C^{2|I|} \delta^{2|I|}.  \end{equation}
Summing over the intervals $[t_{j_i},t_{j'_i}]$ of length $\delta$ for each $i_l$ ($l=1,\dots,p$), we have
\begin{equation} \pi_\sigma(\nu_n)(\pi_\sigma^{-1} \left( \bigcap_{i \in I}
\{\xi:\,\xi_i \in G_\delta^c\} \right) < C^{2|I|} \delta^{|I|}. \end{equation}
It follows that 
\begin{eqnarray} \lefteqn{\pi_\sigma(\nu_n) \left(\pi_\sigma^{-1} \big( \bigcap_{i \in I} 
\{\xi:\,\xi_i \in K(\delta)^c\} \big) \right)} \non \\ &=& \sum_{n_1,\dots,n_{|I|}=1}^\infty \pi_\sigma(\nu_n) \left(\pi_\sigma^{-1} 
\big( \bigcap_{i \in I} \{\xi:\,\xi_i \in G_{\delta/n_i^2}^c\} \big) \right) \non \\ 
&\leq & \sum_{n_1,\dots,n_{|I|}=1}^\infty \pi_\sigma(\nu_n) \left( \pi_\sigma^{-1} 
\big( \bigcap_{i \in I} \{\xi:\,\xi_i \in G_{\delta/(\max(n_i)^2)}^c\} \big) \right)  \non \\ 
&\leq & C^{2|I|} \delta^{|I|} \sum_{n_1,\dots,n_{|I|}=1}^\infty \max(n_i)^{-2|I|} 
\leq \left(\frac{\pi^2 C^2 \delta}{6}\right)^{|I|}. \end{eqnarray}
Finally, we replace $\delta$ by $6 \delta/(\pi^2 C^2)$ as before. \qed

\begin{prop} \label{QLD-Prop3} The following large-deviation upper bound holds.
\begin{equation} \limsup_{n \to \infty} \frac{1}{n} \ln \Tr e^{n[f(X^{(n)}) + g(Y^{(n)})]} 
\leq  \sup_{x \in S_1} \sup_{y \in S_2} [f(x) + g(y) - I(x,y)] . \end{equation}
\end{prop}

\Pf We  set
\begin{equation}  \label{Hndef} H_n = n g(Y^{(n)}). \end{equation} 
 The corresponding $\Gamma_n$ is given by 
 \begin{eqnarray} \lefteqn{(\Gamma_n)_{\ul{k},\ul{k}'} = } \non \\ && \begin{cases}  
 n \left| \sum_{p=1}^d  \frac{c_p}{n^p} 
\sum_{i_1, \dots, i_p = 1}^n \prod_{l=1}^p (Y_{i_l})_{k_{i_l},k'_{i_l}} \prod_{j\neq i_l} \delta_{k_j,k'_j} \right|,
&\text{if $\ul{k} \neq \ul{k}'$;} \\ - n \sum_{\ul{k}'' \neq \ul{k}}  
\left| \sum_{p=1}^d \frac{c_p}{n^p} 
\sum_{i_1, \dots, i_p = 1}^n \prod_{l=1}^p (Y_{i_l})_{k_{i_l},k''_{i_l}} \prod_{j\neq i_l} \delta_{k_j,k'_j}  \right|,
&\text{if $\ul{k} = \ul{k}'$.} \end{cases} \end{eqnarray}
In particular, 
\begin{eqnarray} \lefteqn{(\Gamma^{(i)}_n)_{\ul{k},\ul{k}'} = } \non \\ && \begin{cases}  
n \left| \sum_{p=1}^d \frac{c_p}{n^p} 
\sum_{i_1, \dots, i_{p}:\,(\exists l)\,i_l = i} \prod_{l=1}^p (Y_{i_l})_{k_{i_l},k''_{i_l}} 
\prod_{j\neq i_l} \delta_{k_j,k'_j} \right|, 
&\text{if $k_i \neq k'_i$;}  \\ 0 &\text{otherwise.} \end{cases} 
\end{eqnarray} 
Therefore, 
\begin{eqnarray} || \Gamma^{(i)}_n|| &\leq & n \sup_{\ul{k} \in \NN_m^n}
\sum_{p=1}^d  \frac{|c_p|}{n^p} 
\sum_{i_1, \dots, i_{p}:\,(\exists l)\,i_l = i} \sum_{k'_i \in \NN_m;\,k'_i \neq k_i}
\non \\ &&\qquad \qquad  \times \prod_{j \in \{i_1,\dots,i_p\}} \sum_{k'_{j}} |(Y_j^{m_j})_{k_{j},k'_{j}} \prod_{l \notin \{1,\dots,i_p\}} \delta_{k_l,k'_l}|
\non \\ &\leq & \sum_{p=1}^d \frac{|c_p|}{n^p} p\,n^{p-1} ||Y||_*^p 
= \sum_{p=1}^d |c_p| \,p  ||Y||_*^p  < +\infty. \end{eqnarray} 
Moreover, $$ || \Gamma_n^{(i,j)} || \leq ||\Gamma_n^{(i)}||\,||\Gamma_n^{(j)}||. $$ 
Indeed, 
$||\Gamma_n^{(i,j)}|\ = O(n^{-1})$ is negligible: it is very unlikely that $\xi_i$ and $\xi_j$ jump 
at the same time. The conditions for Lemma~\ref{QLD-L5-3} are therefore satisfied. 
Similar to Lemma~\ref{QLD-L3-1}, this implies that there exists, for given $L > 0$, 
a compact set $K_L \subset L^2([0,1])$ such that 
\begin{equation} \label{Avtight} 
\limsup_{n \to \infty} \frac{1}{n} \ln |\mu_n| \left( \{\xi \in L^2([0,1],\RR^n):\,
\frac{1}{n} \sum_{i=1}^n \tilde{\lambda} \circ \xi_i \in K_L^c \} \right) \leq -L. 
\end{equation} 
To prove this, note that by Lemma~\ref{QLD-L5-3} there exists a compact set $K(\delta)$ such that 
\begin{equation} \nu_n (\{(\xi_1,\dots,\xi_n) \in D([0,1],\NN_m^n):\, (\forall i \in I) \xi_i \in K(\delta)^c\}) < \delta^{|I|}. \end{equation} for $I \subset \{1,\dots,n\}$. Let 
$\tilde{K}(\delta) = \tilde{\lambda}(K(\delta))$. Then 
\begin{equation} \tilde{\nu}_n (\{(\eta_1,\dots,\eta_n) \in D([0,1],\NN_m^n):\, (\forall i \in I) \eta_i \in \tilde{K}(\delta)^c\}) < \delta^{|I|}. \end{equation}
where $\tilde{\nu}_n = \tilde{\lambda}(\nu_n)$. Then we define 
\begin{equation} G(\eps, \delta) = \{ \alpha \in {\cal M}_1^+(L^2([0,1])):\, \alpha(\tilde{K}(\delta)^c) \leq \eps \}, \end{equation}
where $\eps \in (0,\half]$ and $\delta < \eps e^{-2/\eps e}$. Then 
\begin{eqnarray} \lefteqn{\tilde{\nu}_n \left(\{ \eta:\, \frac{1}{n} \sum_{i=1}^n \delta_{\eta_i} \in G(\eps,\delta)^c \} \right) } \non \\ &=& 
\tilde{\nu}_n \left(\{ \eta:\, \frac{1}{n} \#\{i:\,\eta_i \in \tilde{K}(\delta)^c\} > \eps \} \right) \non \\ 
&\leq & \sum_{\substack{I \subset \{1,\dots,n\}: \\ |I| \geq n\eps}} 
\tilde{\nu}_n (\{\eta \in D([0,1],\NN_m^n):\, (\forall i \in I) \eta_i \in \tilde{K}(\delta)^c\})  \non \\ 
&\leq & \sum_{p=[n\eps]}^n {n \choose p} \delta^p. \end{eqnarray}
Taking logarithms, we can use Stirling's formula to find that 
\begin{equation} \limsup_{n \to \infty} \frac{1}{n} \ln \tilde{\nu}_n \left(\{ \eta:\, \frac{1}{n} \sum_{i=1}^n \delta_{\eta_i} \in G(\eps,\delta)^c \} \right)  \leq \eps \ln \frac{\delta}{\eps}.  
\end{equation} This is the analogue of equation (\ref{Gbnd}).
As in the proof of Lemma~\ref{QLD-L3-1} this implies that there exists a compact set $K_L \subset L^2([0,1])$ such that 
\begin{equation} \limsup_{n \to \infty} \frac{1}{n} \ln \tilde{\nu}_n \left( \{\eta \in L^2([0,1],\RR^n):\,
\frac{1}{n} \sum_{i=1}^n \eta_i \in K_L^c \} \right) \leq -L. 
\end{equation} 
Finally, we note that $|\mu_n|(\tilde{\lambda}^{-1}(A)) = \nu_n(\tilde{\lambda}^{-1}(A)) 
\Tr e^{\Gamma_n}$, where 
$$ || \Gamma_n|| \leq n \sum_{p=1}^d c_p ||Y||_*^p. $$
Therefore, we can replace $K_L$ by $\tilde{K}_L = K_{L'}$ with 
$L' = L + \sum_{p=1}^d c_p ||Y||_*^p$ to obtain 
\begin{equation} \limsup_{n \to \infty} \frac{1}{n} \ln |\mu_n| \left( \{\xi \in L^2([0,1],\RR^n):\,
\frac{1}{n} \sum_{i=1}^n \tilde{\lambda} \circ \xi_i \in \tilde{K}_L^c \} \right) \leq -L. 
\end{equation} 

As in the proof of Proposition~\ref{QLD-Prop1}, this implies that it suffices to show that 
\begin{align} &\limsup_{n \to \infty} \frac{1}{n} \ln \int_{A_n^{-1}(K_L)} \exp \left\{ n \int_0^1 
f \big(\frac{1}{n} \sum_{k=1}^n \tilde{\lambda}(\xi_k(t)) \big) dt \right\} \non \\  
&\qquad \qquad \times \exp \left\{- \ \int_0^1 H_D(\xi) \, dt \right\} 
\mu_n(d\xi) \non \\ &\leq \sup_{(x,y) \in S_1 \times S_2} [f(x) + g(y) - I(x,y)]. 
\end{align}
Let $\kappa_n$ be the image measure $\kappa_n = (A_n \circ \tilde{\lambda})(\mu_n)$. Then we can write this again as 
\begin{align} &\limsup_{n \to \infty} \frac{1}{n} \ln \int_{K_L} \exp \left\{ n \int_0^1 
f \big(\eta(t)) \big) dt \right\} \kappa_n(d\eta) \non \\ 
&\leq \sup_{(x,y) \in S_1 \times S_2} [f(x) + g(y) - I(x,y)].  \end{align}
Introducing the Haar basis again, we can replace $f(\eta(t))$ by $f(\pi_N(\eta(t)))$ as before
and write
\begin{align} &\int_{K_L} \exp \left\{ n \int_0^1 
f \big(\eta(t)) \big) dt \right\} \kappa_n(d\eta) \non \\ 
&= \sum_{\ul{j}_1, \dots, \ul{j}_N \in \NN_m^n} \prod_{k=1}^N e^{n f(\eta(\ul{j}_k))/N} 
\kappa_n(\{\eta:\, \eta(k/N) = \eta(\ul{j}_k) \, (k=1,\dots,N)\}). \end{align}
(Here, by abuse of notation, $\eta(\ul{j}_k) = \frac{1}{n} \sum_{i=1}^n \tilde{\lambda}(j_{k,i})$.)
The $\kappa_n$ measure equals
$$ \kappa_n(\{\eta:\, \eta(k/N) = \eta(\ul{j}_k) \, (k=1,\dots,N)\}) = 
\Tr \left[ P^1_{\ul{j}_N} e^{- H_n/N} \dots P^1_{\ul{j}_1} e^{- H_n/N} \right] $$ and therefore
\begin{align}&\int_{K_L} \exp \left\{ n \int_0^1 
f \big(\eta(t)) \big) dt \right\} \kappa_n(d\eta)  \non \\ &\leq  \sum_{i_1,\dots,i_N=1}^{r_1} 
\prod_{k=1}^N e^{n [f(x_{i_k}) + \eps]/N} 
\left| \Tr \left[ P_{i_N} e^{- H_n/N} \dots P_{i_1} e^{- H_n/N} \right] \right|. \end{align}
Expanding $H_n = n g(Y^{(n)})$ into spectral projections, we get
\begin{align}&\int_{K_L} \exp \left\{ n \int_0^1 
f (\eta(t))  dt \right\} \kappa_n(d\eta)  \non \\ &\leq  \sum_{i_1,\dots,i_N=1}^{r_1} 
\prod_{k=1}^N e^{n [f(x_{i_k}) + \eps]/N} e^{n [g(y_{j_k}) + \eps]/N}
\left| \Tr \left[ P_{i_N} Q_{j_N} \dots P_{i_1} Q_{j_1} \right] \right| \non \\ 
&\leq \sum_{i_1,\dots,i_N=1}^{r_1} 
\prod_{k=1}^N e^{n [f(x_{i_k}) + \eps]/N} e^{n [g(y_{j_k}) + \eps]/N} \non \\ 
&\qquad\qquad  \times \exp \left[ -n \sum_{i=1}^{r_1} \sum_{j=1}^{r_2} \gamma_{i,j} 
\inf_{\substack{x \in [x_{i-1},x_i] \\ y \in [y_{i-1},y_i]}} I(x,y) + n \eta\right] \non \\ 
&\leq \sum_{\substack{\{N_{i,j}\}_{i,j=1}^{r_1,r_2}\\ \sum N_{i,j} = N}} \frac{N!}{\prod_{i,j} N_{i,j}!}
\prod_{i=1}^{r_1} \prod_{j=1}^{r_2} e^{n \gamma_{i,j} [f(x_i) + g(y_j) + 2\eps]} \non \\ 
&\qquad \qquad \times \exp \left[ - n\inf_{\substack{x \in [x_{i-1},x_i] \\ y \in [y_{i-1},y_i]}} 
I(x,y) + n \eta \right]
\end{align}
Since $N$ is finite, the limit yields 
\begin{align} &\limsup_{n \to \infty} \frac{1}{n} \ln \int_{K_L} \exp \left\{ n \int_0^1 
f (\eta(t)) \, dt \right\} \kappa_n(d\eta)  \non \\ &\leq
\sup_{(x,y) \in S_1 \times S_2} \,[f(x) + g( y) - I(x,y)] + \eta + 2\eps. \end{align}
This proves the LD upper bound. \qed

\setcounter{equation}{0}

\section{Multivariable generalization}

We would like to generalize Proposition~\ref{QLD-Prop3} further to several variables, that is,  
to an arbitrary number of operators $X_1,\dots,X_q$. However, Lemma~\ref{QLD-L5-1} does not 
extend to a product of more than two projections. Instead, we have to iterate the procedure 
in the proof of Proposition~\ref{QLD-Prop3}.

\begin{prop} \label{QLD-Prop4} Let $X_1, \dots, X_q$ ($q \in \NN$) be self-adjoint matrices in $\cal M$, and let $f_1, \dots, f_q$ be continuous functions $f_j: {\rm co}(\sigma(X_j)) \to \RR$ 
($j=1,\dots, q$).
Define the cumulant generating function $C: \RR^q \to \RR$ by 
\begin{equation} \label{CGFq} C(t_1,\dots,t_q) = \ln \Tr e^{t_1 X_1 + \dots + t_q X_q}, \end{equation} 
and let $I: \RR^q \to [0,+\infty]$ be the Legendre transform.
Then the following LD upper bound holds.
\begin{align} &\limsup_{n \to \infty} \frac{1}{n} \ln \Tr e^{n[f_1(X_1^{(n)}) + \dots + f_1(X_q^{(n)})]} 
\non \\ &\qquad \leq  \sup_{\substack{(x_1,\dots,x_q) \in \RR^q: \\
\forall j:\,x_j \in {\rm co}(\sigma(X_j))}} [f_1(x_1) + \dots + f_q(x_q) - I(x_1\dots,x_q)]. 
\end{align}
\end{prop}

\Pf
Note Lemma~\ref{QLD-L5-1} does not generalize. Instead, we use the Trotter formula one factor 
at a time. 
Consider the case $q=3$. The partition function equals
\begin{equation} {\cal Z} = \Tr \big[e^{n[f(X^{(n)}) + g(Y^{(n)}) + h(Z^{(n)})]}\big]. \end{equation}
We introduce first the Hamiltonian $H_n = n[g(Y^{(n)}) + h(Z^{(n)})]$. 
Then there exists a complex-valued measure $\kappa_n$ on $L^2([0,1])$ such that 
\begin{equation} {\cal Z} = \int \exp 
\left[ n \int_0^1 f \left( \frac{1}{n} \sum_{i=1}^n \eta_i(t) \right) dt \right] \kappa_n(d\eta). 
\end{equation} 
As in the proof of Proposition~\ref{QLD-Prop3}, the paths $\eta_i(t)$ jump rarely 
(see Lemma~\ref{QLD-L5-3}), and this implies 
that, given $L > 0$, there exists a compact set $K_L$ such that 
$$ \limsup_{n \to \infty} \frac{1}{n} \ln |\mu_n| \left( \{ \eta \in L^2([0,1],\RR^n):\, 
\frac{1}{n} \sum_{i=1}^n \tilde{\lambda} \circ \xi_i \in K_L^c \} \right) \leq -L. $$
(This is analogous to (\ref{Avtight}).) This means again that we can replace $f$ by $f \circ \pi_{N_1}$
for large enough $N_1$. Now taking $N = M N_1$ to be a multiple of $N_1$, we have 
\begin{eqnarray} {\cal Z} &=& \lim_{M \to \infty} \Tr \left[ \big(e^{n (f \circ \pi_N)(X^{(n)})/MN_1} 
e^{n [g(Y^{(n)}) + h(Z^{(n)})]/MN_1} \big)^{MN_1} \right] \non \\
&=&   \sum_{i_1,\dots,i_{N_1}=1}^{r_1} \prod_{k=1}^{N_1} e^{n f(x_{i_k})/N_1} 
\non \\ && \qquad  \times \lim_{M \to \infty} 
\Tr \left[ \prod_{k=1}^{N_1} \left(  \big( P_{i_k} e^{n (g(Y^{(n)}) + h(Z^{(n)}))/MN_1} \big)^{M}\right) \right] \non \\ 
&\leq & \sum_{i_1,\dots,i_{N_1}=1}^{r_1}  \prod_{k=1}^{N_1} e^{n f(x_{i_k})/N_1} 
\non \\ && \qquad  \times \prod_{k=1}^{N_1} \lim_{M \to \infty} 
\left\{ \Tr \left[ \left(  P_{i_k} e^{n (g(Y^{(n)}) + h(Z^{(n)}))/MN_1} 
\right)^{MN_1} \right] \right\}^{1/N_1}.  \end{eqnarray} 
Now, given any $t_1 \in \RR$, we define $a_i(t_1) = \begin{cases} x_i &\text{if $t_1 < 0$;} \\ x_{i-1} &\text{if $t_1 > 0$.} \end{cases}$. \\ 
Then the eigenprojection corresponding to $X$ is bounded by \\
$P^{(1)}_i \leq e^{n t_{1,i} (X^{(n)}-a_i(t_{1,i}))/N}$. Inserting this, we have
\begin{eqnarray} {\cal Z}  &\leq & \sum_{i_1,\dots,i_{N_1}=1}^{r_1} 
\prod_{k=1}^{N_1} e^{n f(x_{i_k})/N_1} \non \\ && \times \prod_{k=1}^{N_1}
\left\{ \Tr \left[  \left( e^{n t_{1,i_k} (X^{(n)}-a_{i_k}(t_{1,i_k}))/MN_1} 
e^{n (g(Y^{(n)}) + h(Z^{(n)}))/MN_1} \right)^{MN_1} \right]\right\}^{1/N_1}. \end{eqnarray} 
Next, we use the Lie-Trotter theorem to obtain
\begin{eqnarray} {\cal Z} &\leq& \sum_{i_1,\dots,i_{N_1}} \prod_{k=1}^{N_1} 
e^{n f(x_{i_k}) /N_1} \non \\ && 
\times \prod_{k=1}^{N_1} \left\{ e^{-n t_{1,i_k}  a_{i_k}(t_{1,i_k})}
\Tr \left[ e^{n \big(t_{1,i_k} X^{(n)} + g(Y^{(n)}) + h(Z^{(n)}) \big)}   \right] \right\}^{1/N_1}\!\!.
\end{eqnarray}

We now repeat this process with the new Hamiltonian 
$$ H'_n = n \big(t_{1,i_k} X^{(n)} + h(Z^{(n)}) \big). $$
Expanding now according to the eigenstates of $Y^{(n)}$, we have in an analogous fashion,
\begin{eqnarray} \lefteqn{\Tr \left[ e^{n \big(t_{1,i_k}) X^{(n)} + g(Y^{(n)}) + h(Z^{(n)}) \big)}   \right]} \non \\ &\leq & \sum_{j_1,\dots,j_{N_2}=1}^{r_2} 
\prod_{k=1}^{N_2} e^{n g(y_{j_k})/N_2} \non \\ && \qquad \times
\prod_{k=1}^{N_2} \lim_{M \to \infty} \left\{ \Tr \left[ \left( Q_{j_k} e^{n( (t_{1,i_k}) X^{(n)}  + h(Z^{(n)}))/MN_2} \right)^{MN_2} \right]  \right\}^{1/N_2} \end{eqnarray}
Again, we define for any $t_2 \in \RR$, $b_j(t_2) = \begin{cases} y_j &\text{if $t_2 < 0$;} \\ y_{j-1} &\text{if $t_2 > 0$.} \end{cases}$. \\ 
Then $P^{(2)}_j \leq e^{n t_{2,j} (Y^{(n)}-b_j(t_{2,j}))/N}$.
Inserting, and taking the limit $M \to \infty$, we get
\begin{eqnarray} { \cal Z} &\leq&
\sum_{i_1,\dots,i_{N_1}} \sum_{j_1,\dots,j_{N_2}} \prod_{k=1}^{N_1}    e^{n f(x_{i_{k_1}})/N_1}  \prod_{k_2=1}^{N_2} e^{n g(y_{j_{k_2}})/N_2} \non \\ &&  \times \prod_{k_1=1}^{N_1} 
e^{-n  t_{1,i_{k_1}} a_{i_{k_1}}(t_{1,i_{k_1}})/N_1} 
\prod_{k_2=1}^{N_2}  e^{-n t_{2,j_{k_2}}  b_{j_{k_2}}(t_{2,j_{k_2}})/N_2 }
\non \\ && \times \prod_{k_1=1}^{N_1} \prod_{k_2=1}^{N_2}  \left\{ 
\Tr \left[ e^{n \big(t_{1,i_{k_1}} X^{(n)} + t_{2,j_{k_2}}  Y^{(n)} + h(Z^{(n)}) \big)}   
\right] \right\}^{1/N_1 N_2}. \end{eqnarray}
Repeating this procedure  once more, we obtain 
\begin{eqnarray} {\cal Z} &\leq& \sum_{i_1,\dots,i_{N_1}} \sum_{j_1,\dots,j_{N_2}} 
\sum_{l_1,\dots,l_{N_3}} \non \\ && \qquad \times
\prod_{k_1=1}^{N_1}   e^{n f(x_{i_{k_1}})/N_1} \prod_{k_2=1}^{N_2} e^{n g(y_{j_{k_2}})/N_2}
\prod_{k_3=1}^{N_3}   e^{n h(z_{l_{k_3}})/N_3} \non \\ &&  \times 
\prod_{k_1=1}^{N_1} e^{-n t_{1,i_{k_1}} a_{i_{k_1}}(t_{1,i_{k_1}})/N_1} 
\prod_{k_2=1}^{N_2}  e^{-n t_{2,j_{k_2}}  b_{j_{k_2}}(t_{2,j_{k_2}})/N_2} 
\prod_{k_3=1}^{N_3}  e^{-n t_{3,l_{k_3}}  c_{l_{k_3}}(t_{3,l_{k_3}})/N_3}
\non \\ && \times \prod_{k_1=1}^{N_1} \prod_{k_2=1}^{N_2}   \prod_{k_3=1}^{N_3}  \left\{
\Tr \left[ e^{n \big(t_{1,i_{k_1}} X^{(n)} + t_{2,j_{k_2}} Y^{(n)} + t_{3,l_{k_3}}  Z^{(n)} \big)}   
\right] \right\}^{1/N_1 N_2 N_3}. \end{eqnarray} 
The latter trace is just the exponential of the cumulant generating function, so that 
\begin{eqnarray} {\cal Z} &\leq& \sum_{i_1,\dots,i_{N_1}} \sum_{j_1,\dots,j_{N_2}} 
\sum_{l_1,\dots,l_{N_3}}  \prod_{k_1=1}^{N_1}   e^{n f(x_{i_{k_1}})/N_1} 
\non \\ && \times \prod_{k_2=1}^{N_2} e^{n g(y_{j_{k_2}})/N_2}
\prod_{k_3=1}^{N_3}   e^{n h(z_{l_{k_3}})/N_3} 
\prod_{k_1=1}^{N_1} e^{-n t_{1,i_{k_1}} a_{i_{k_1}}(t_{1,i_{k_1}})/N_1} \non \\ &&  \times 
\prod_{k_2=1}^{N_2}  e^{-n t_{2,j_{k_2}}  b_{j_{k_2}}(t_{2,j_{k_2}})/N_2} 
\prod_{k_3=1}^{N_3}  e^{-n t_{3,l_{k_3}}  c_{l_{k_3}}(t_{3,l_{k_3}})/N_3}
\non \\ && \times \prod_{k_1=1}^{N_1} \prod_{k_2=1}^{N_2}   \prod_{k_3=1}^{N_3} 
e^{n C(t_{1,i_{k_1}}, t_{2,j_{k_2}}, t_{3,l_{k_3}})/N_1 N_2 N_3} . \end{eqnarray}

Now assume that the supremum  in 
$I(x_0,y_0,z_0) = \sup_{(t_1,t_2,t_3) \in \RR^3} [t_1 x_0 + t_2 y_0 + t_3 z_0) - C(t_1,t_2,t_3)]$
is attained at $(t_1(x_0,y_0,z_0), t_2(x_0,y_0,z_0), t_3(x_0,y_0,z_0))$.
Setting $t_{1,i_k} = t_1(x_{i_k},y_{j_k},z_{l_k})$ etc. we conclude that 
\begin{eqnarray} {\cal Z} &\leq& \sum_{i_1,\dots,i_{N_1}} \sum_{j_1,\dots,j_{N_2}} 
\sum_{l_1,\dots,l_{N_3}} \non \\ && \qquad \times
\prod_{k_1=1}^{N_1}   e^{n f(x_{i_{k_1}})/N_1} \prod_{k_2=1}^{N_2} e^{n g(y_{j_{k_2}})/N_2}
\prod_{k_3=1}^{N_3}   e^{n h(z_{l_{k_3}})/N_3} \non \\ &&  \times 
\prod_{k_1=1}^{N_1} \prod_{k_2=1}^{N_2}   \prod_{k_3=1}^{N_3} 
\exp \bigg[ - \frac{n}{N_1 N_2 N_3} \inf_{\substack{x \in [x_{i_{k_1}-1}, x_{i_{k_1}}] \\ y \in
[y_{j_{k_2}-1}, y_{j_{k_2}}] \\ z \in [z_{l_{k_3}-1}, z_{l_{k_3}}]}} I(x,y,z) \bigg]. \end{eqnarray}
As before, this can be  written as 
\begin{eqnarray} {\cal Z} &\leq& \sum_{\substack{M_1,\dots,M_{r_1} \geq 0 \\ \sum M_i = N_1}}
\frac{N_1!}{M_1! \dots M_{r_1}!} 
\sum_{\substack{M'_1,\dots,M'_{r_2} \geq 0 \\ \sum M'_i = N_2}} \frac{N_2!}{M'_1! \dots M'_{r_2}!} 
\\ && \qquad \times
\sum_{\substack{M''_1,\dots,M''_{r_3} \geq 0 \\ \sum M''_i = N_3}} \frac{N_3!}{M''_1! \dots M''_{r_3}!} 
\prod_{i=1}^{r_1}   \prod_{j=1}^{r_2} \prod_{l=1}^{r_3} \bigg\{
e^{n \gamma_{i,j,l} [f(x_i) +  g(y_j)+ h(z_l)]} \non \\ &&  \times 
\exp \big[ - n \gamma_{i,j,l}  \inf_{(x,y,z) \in [x_{i-1}, x_{i}] \times
[y_{j-1}, y_{j}] \times [z_{l-1}, z_{l}]} I(x,y,z) \big] \bigg\}. \end{eqnarray}
Since 
$$ \sum_{\substack{M_1,\dots,M_{r_1} \geq 0 \\ \sum M_i = N_1}}
\frac{N_1!}{M_1! \dots M_{r_1}!} = (r_1)^{N_1}, \mbox{ etc.} $$
and $N_1$, $N_2$ and $N_3$ are independent of $n$,
we can take the logarithm and divide by $n$ to get
\begin{eqnarray} \lefteqn{\limsup_{n \to \infty} \frac{1}{n} \ln 
\Tr [e^{n(f(X^{(n)}) + g(Y^{(n)}) + h(Z^{(n)}))}]}  \non \\ &\leq&
\sup_{(x,y,z) \in S_1 \times S_2 \times S_3}  [f(x) + g(y) + h(z) - I(x,y,z)] + 3\eps. \end{eqnarray}
assuming that $f$, $g$ and $h$ do not vary by more than $\eps$ over the intervals 
$[x_{i_{k_1}-1}, x_{i_{k_1}}]$,  $[y_{j_{k_2}-1}, y_{j_{k_2}}]$ and $ [z_{l_{k_3}-1}, z_{l_{k_3}}] $
respectively. Taking $\eps \to 0$, the upper bound follows. It is clear that this procedure can be 
repeated to obtain for any finite $q$, the upper bound
\begin{eqnarray} \lefteqn{\limsup_{n \to \infty} \frac{1}{n} \ln 
\Tr [e^{n [f_1(X_1^{(n)}) + \dots + f_q(X_q^{(n)})]}]}  \non \\ &\leq&
\sup_{(x_1,\dots,x_q) \in S_1 \times \dots \times S_q}  [f_1(x_1) + \dots + f_q(x_q) - I(x_1, \dots, x_q)].
\end{eqnarray} 
\qed

To prove the lower bound, we need to generalize Corollary~\ref{QLD-cor2}. 
First note that Lemma~\ref{QLD-L1-1} can be generalized to 
\begin{lemma} \label{QLD-L6-1} The cumulant generating function $C(t_1,\dots,t_q)$ defined by (\ref{CGFq}) is a jointly convex function and its derivatives are given by 
\begin{equation} \frac{\partial C(t_1,\dots,t_q)}{\partial t_k} = \frac{\Tr [X_k e^{t_1 X_1 + \dots + t_q X_q}]}{\Tr [e^{t_1 X_1 + \dots + t_q X_q}]}. \end{equation}
\end{lemma}
This is proved in the same way as Lemma~\ref{QLD-L1-1}.

Lemma~\ref{QLD-L4-2} becomes
\begin{lemma} \label{QLD-L6-2} Let $X_1,\dots,X_q \in {\cal M}$ be hermitian matrices. Then
\begin{equation} \ln \Tr e^{t_1 X_1 + \dots + t_1 X_q} \geq t_1 \varphi(X_1) + \dots + t_q \varphi(X_q) 
-S(\varphi\,||\,\tau) + \ln m \end{equation} for every state $\varphi \in {\cal S}$, where $\tau$ is the tracial state $\tau(X) = \frac{1}{m} \Tr(X)$. Moreover, equality holds only when $\varphi = \omega_{t_1,\dots,t_q}$, where
\begin{equation} \omega_{t_1,\dots,t_q}(A) = 
\frac{\Tr [A \,e^{t_1 X_1 + \dots + t_q X_q}]}{\Tr [e^{t_1 X_1 + \dots + t_q X_q}]}. \end{equation} 
\end{lemma}

The generalization of Lemma~\ref{QLD-L4-3} is
\begin{lemma} \label{QLD-L6-3} Let $\tau$ be the tracial state on $\cal M$. 
For every state $\varphi \in {\cal S}$, and any set of hermitian $X_1,\dots, X_q \in {\cal M}$,
$$ S(\varphi\,||\, \tau) \geq I(\phi(X_1),\dots,\varphi(X_q)) + \ln m. $$ 
Moreover, if $S(\varphi\,||\, \tau) =I(\phi(X_1),\dots,\varphi(X_q)) + \ln m < +\infty$ 
and there exist $t_1,\dots,t_q \in \RR$ such that 
$I(\varphi(X_1), \dots, \varphi(X_q)) = t_1 \varphi(X_1) + \dots + t_q \varphi(X_q) - C(t_1,\dots,t_q)$
then  $\varphi = \omega_{t_1,\dots,t_q}$. 
\end{lemma}

\Pf By Lemma~\ref{QLD-L6-2}, for every set of hermitian matrices $X_1, \dots, X_q \in {\cal M}$, 
and any $t_1, \dots, t_q \in \RR$, 
$$ S(\varphi\,||\, \tau) \geq t_1 \varphi(X_1) + \dots + t_q \varphi(X_q) - C(t_1, \dots, t_q) + \ln m, $$
and maximizing over $t_1,\dots, t_q$, 
$$ S(\varphi\,||\, \tau) \geq I(\varphi(X_1), \dots, \varphi(X_q)) + \ln m. $$
Moreover, if $\varphi = \omega_{t_1,\dots,t_q}$ then equality holds. 

Conversely, suppose that $S(\varphi\,||\, \tau) =I(\varphi(X_1),\dots,\varphi(X_q)) + \ln m < +\infty$. 
If there exist $t_1,\dots, t_q \in \RR$ such that $I(\phi(X_1),\dots,\varphi(X_q))  =  t_1\varphi(X_1) + \dots + t_q\varphi(X_q) - C(t_1,\dots, t_q)$ then by the uniqueness in Lemma~\ref{QLD-L6-2}, 
$\varphi = \omega_{t_1,\dots,t_q}$.  \qed

%Otherwise, there exists a sequence $(\ul{t}_n)_{n \in \NN}$ in $\RR^q$ such that $|\ul{t}_n| \to \infty$ %and $I(\varphi(X_1),\dots,\varphi(X_q)) = 
%\lim_{n \to \infty} [t_1 \varphi(X_1) + \dots + t_q \varphi(X_q) - C(\ul{t})].$

As a consequence we have
\begin{cor} \label{QLD-cor3} For any continuous function $F: S_1 \times \dots \times S_q \to \RR$ the following identity holds. 
\begin{equation} \sup_{\ul{u} \in S_1 \times \dots \times S_q} [F(\ul{u}) - I(\ul{u})] = 
\sup_{\varphi \in {\cal S}} [F(\varphi(X_1), \dots, \varphi(X_q)) - S(\varphi\,||\,\tau)] + \ln m. 
\end{equation}
\end{cor}

\Pf Clearly, 
\begin{eqnarray*} \lefteqn{\sup_{\varphi \in {\cal S}}\, [F(\varphi(X_1), \dots, \varphi(X_q)) 
- S(\varphi\,||\,\tau)] + \ln m} \\  
&\leq & \sup_{\varphi \in {\cal S}}\, [F(\varphi(X_1), \dots, \varphi(X_q)) - 
I(\varphi(X_1), \dots, \varphi(X_q))] \\ &\leq & \sup_{\ul{u} \in S_1 \times \dots \times S_q} 
[F(\ul{u}) - I(\ul{u})] \end{eqnarray*}
since $I(\ul{u}) = +\infty$ if $\ul{u} \notin S_1 \times \dots \times S_q$.
On the other hand, if there exists $\ul{t} \in \RR^q$ such that $I(\ul{u}) = \langle \ul{t}, \ul{u} \rangle - C(\ul{t})$ then we put $\varphi = \omega_{\ul{t}}$. By Lemma~\ref{QLD-L6-3},
$S(\omega_{\ul{t}}\,||\, \tau) = t_1 \varphi(X_1) + \dots + t_q \varphi(X_q) - C(\ul{t}) - \ln m$ 
and $\varphi(X_k) = u_k$ since $u_k = \partial C(\ul{t})/\partial t_k$. Therefore $S(\varphi\,||\, \tau) = I(\ul{u}) + \ln m$ and $F(\ul{u}) - I(\ul{u}) = F(\varphi(X_1), \dots, \varphi(X_q)) - S(\varphi\,||\,\tau)
+ \ln m.$
Finally, note that $|\nabla I(\ul{u})| \to \infty$ as $\ul{u}$ tends to the boundary of ${\cal D}(I)$. 
Let $(\ul{u}_n)_{n \in \NN}$ be a sequence in the relative interior of ${\cal D}(I)$ such that 
$I(\ul{u}_n) \to I(\ul{u})$. Then, for large enough $n$, $I(\ul{u}_n) < I(\ul{u})$ and given $\eps > 0$,
$|F(\ul{u}_n) - F(\ul{u})| < \eps$. But then 
\begin{eqnarray*} F(\ul{u}) - I(\ul{u}) &\leq & F(\ul{u}) - I(\ul{u}_n) \\ &=& 
F(\ul{u}) - S(\omega_{\ul{t}_n}\,||\, \tau) + \ln m \\ &\leq & 
F(\ul{u}) - F(\ul{u}_n) \\ && \qquad + \sup_{\varphi \in {\cal S}}\, 
[F(\varphi(X_1), \dots, \varphi(X_q)) - S(\varphi\,||\,\tau)] + \ln m \\ &\leq & 
\sup_{\varphi \in {\cal S}}\,[F(\varphi(X_1), \dots, \varphi(X_q)) - S(\varphi\,||\,\tau)] + \ln m + \eps. \end{eqnarray*}
Taking $\eps \to 0$ it follows that the inequality also holds for $\ul{u} \in \partial {\cal D}(I)$. \qed

This corollary allows us to prove the LD lower bound in the same way as Proposition~\ref{QLD-Prop2}.

\begin{theorem} \label{QLD-Thm2} Let $X_1, \dots, X_q$ ($q \in \NN$) be self-adjoint matrices 
in $\cal M$, and let $f_1, \dots, f_q$ be continuous functions $f_j: {\rm co}(\sigma(X_j)) \to \RR$ 
($j=1,\dots, q$).
Define the cumulant generating function $C: \RR^q \to \RR$ by 
\begin{equation}  C(t_1,\dots,t_q) = \ln \Tr e^{t_1 X_1 + \dots + t_q X_q}, \end{equation} 
and let $I: \RR^q \to [0,+\infty]$ be the Legendre transform.
Then the following identity holds.
\begin{align} &\lim_{n \to \infty} \frac{1}{n} \ln \Tr e^{n[f_1(X_1^{(n)}) + \dots + f_1(X_q^{(n)})]} 
\non \\ &\qquad =  \sup_{\substack{(x_1,\dots,x_q) \in \RR^q: \\
\forall j:\,x_j \in {\rm co}(\sigma(X_j))}} [f_1(x_1) + \dots + f_q(x_q) - I(x_1\dots,x_q)]. 
\end{align}
\end{theorem}

\Pf By Proposition~\ref{QLD-Prop4}, 
\begin{align} &\limsup_{n \to \infty} \frac{1}{n} \ln \Tr e^{n[f_1(X_1^{(n)}) + \dots + f_1(X_q^{(n)})]} 
\non \\ &\qquad \leq \sup_{\substack{(x_1,\dots,x_q) \in \RR^q: \\
\forall j:\,x_j \in {\rm co}(\sigma(X_j))}} [f_1(x_1) + \dots + f_q(x_q) - I(x_1\dots,x_q)]. 
\end{align}
To prove the lower bound 
\begin{align} &\liminf_{n \to \infty} \frac{1}{n} \ln \Tr e^{n[f_1(X_1^{(n)}) + \dots + f_1(X_q^{(n)})]} 
\non \\ &\qquad \geq  \sup_{\substack{(x_1,\dots,x_q) \in \RR^q: \\
\forall j:\,x_j \in {\rm co}(\sigma(X_j))}} [f_1(x_1) + \dots + f_q(x_q) - I(x_1\dots,x_q)], 
\end{align}
we approximate each of the functions $f_k$ ($k=1,\dots,q$) by polynomials as in the proof of Proposition~\ref{QLD-Prop2}, and subsequently by expressions of the form (\ref{gapprox}):
\begin{equation} f_k(X_{k,1}, \dots, X_{k,n}) = \sum_{l=1}^{d_k} c_{k,l} \frac{l!}{n^l} \sum_{i_1 < \dots < i_l} X_{k,i_1} \dots X_{k,i_l}. \end{equation}
Then, by Lemma~\ref{QLD-L4-2} with $\cal M$ replaced by ${\cal M}_n$, $tX$ by $n \sum_{k=1}^q f_k(\ul{X}_k)$, and with $H=0$, 
\begin{eqnarray} \lefteqn{\frac{1}{n} \ln \Tr e^{n \sum_{k=1}^q f_k(X_{k,1}, \dots, X_{k,n})} } \non
\\ &\geq & \frac{1}{n} \sup_{\varphi \in {\cal S}_n} \big\{ n \sum_{k=1}^q \varphi(f_k(X_{k,1}, \dots, X_{k,n})) - S(\varphi\,|| \,\tau) \big\} + \ln m. \end{eqnarray}
Inserting product states $\varphi = \omega^{\otimes n}$ with $\omega \in {\cal S}$, we have
\begin{eqnarray} \lefteqn{\frac{1}{n} \ln \Tr e^{n \sum_{k=1}^q f_k(X_{k,1}, \dots, X_{k,n})} } \non
\\ &\geq & \sup_{\omega \in {\cal S}} \big\{ \sum_{k=1}^q  \sum_{l=1}^{d_k} c_{k,l} \frac{l!}{n^l} \sum_{1 \leq i_1 < \dots < i_l \leq n} \omega(X_k)^l - S(\omega\,|| \,\tau) \big\} + \ln m. \end{eqnarray}
Taking the limit $n \to \infty$, this simplifies to 
 \begin{eqnarray} \lefteqn{\liminf_{n \to \infty} 
 \frac{1}{n} \ln \Tr e^{n \sum_{k=1}^q f_k(X_{k,1}, \dots, X_{k,n})} } \non
\\ &\geq & \sup_{\omega \in {\cal S}_n} \big\{ \sum_{k=1}^q  \sum_{l=1}^{d_k} c_{k,l}\omega(X_k)^l - S(\omega\,|| \,\tau) \big\} + \ln m \non \\ 
&=& \sup_{\omega \in {\cal S}} \big\{\sum_{k=1}^q f_k(\omega(X)) - S(\omega\,||\,\tau)]\big\} 
+ \ln m \non \\ 
&=& \sup_{\ul{u} \in S_1 \times \dots \times S_q} 
\big\{\sum_{k=1}^q f_k(u_k) - I(\ul{u}) \big\} + \ln m \end{eqnarray}
by Corollary~\ref{QLD-cor3}. \qed

\textbf{Example 1.} We consider again the transverse-field Ising model with Hamiltonian $H_n$ 
given by equation (\ref{TFIsing}). Applying Theorem~\ref{QLD-Thm2} we have to compute the 
cumulant generating function
\begin{equation} C(t_1,t_2) = \ln \Tr e^{t_1 \sigma^z + t_2 \sigma^x} = 
\ln 2 \cosh \sqrt{t_1^2 + t_2^2}. \end{equation}
The corresponding rate function is
$$ I(x,z) = \sup_{(t_1,t_2) \in \RR^2} [t_1 z + t_2 x - C(t_1,t_2)]. $$
It can be evaluated. Differentiating, we have
\begin{eqnarray} z &=& \frac{t_1}{\sqrt{t_1^2 + t_2^2}} \tanh \sqrt{t_1^2 + t_2^2} \non \\
x &=& \frac{t_2}{\sqrt{t_1^2 + t_2^2}} \tanh \sqrt{t_1^2 + t_2^2}. \end{eqnarray}
Therefore 
\begin{equation} \frac{z}{x} = \frac{t_1}{t_2} \mbox{ and } \sqrt{x^2 + z^2} = 
\tanh \sqrt{t_1^2 + t_2^2}. \end{equation} Inserting this we find that 
\begin{equation} I(x,z) = I_0(\sqrt{x^2 + z^2}), \end{equation}
where 
\begin{equation} I_0(u) = \half (1+u) \ln (1+u) + \half (1-u) \ln (1-u) \end{equation} 
is the usual Ising rate function.
Applying Theorem~\ref{QLD-Thm2} it follows that the free energy density is given by 
\begin{equation} \label{TFIsingvar2} f(\beta,h) = \inf_{\substack{(x,z) \in \RR^2 \\ 
x^2 + z^2 \leq 1}}  \big\{ -z^2 - h x + \frac{1}{\beta} I(x,z) \big\}. \end{equation} 
To see that this is equivalent to  the expression (\ref{TFIsingvar}), it suffices to show that 
$$ \inf_{x \in \RR} \,\{ -\beta h x + I(x,z)\} = \tilde{I}(z). $$
But this follows by differentiation with repect to $x$, which gives
$$ \beta h = \frac{\partial I(x,z)}{\partial x} = t_2(x,z). $$ 
Inserting this into the definition of  $I(x,z)$, the identity follows.
However, the expression (\ref{TFIsingvar2}) is more convenient. It can be rewritten by 
setting $x = u \cos \theta$ and $z = u \sin \theta$. The result is
\begin{equation} \label{TFIsingvar3} f(\beta,h) = \inf_{u \in [0,1],\,\theta \in [0,2\pi]} 
\big\{ -u^2 \sin^2 \theta  - h u \cos \theta + \frac{1}{\beta} I_0(u) \big\}. \end{equation} 
This formula was first derived by Ceg\l a, Lewis and Raggio~\cite{CLR1988}.

\textbf{Example 2.} Consider the mean-field Heisenberg model with Hamiltonian
\begin{equation} H_n^{\rm Heis} = - J \frac{1}{n} \sum_{i,j=1}^n (\sigma^x_i \sigma^x_j + 
\sigma^y_i \sigma^y_j  + \Delta \sigma^z_i \sigma^z_j ). \end{equation} 
To compute the free energy density 
\begin{equation}  f_{\rm Heis}(\beta,J) = -\frac{1}{\beta} \lim_{n \to \infty} \frac{1}{n} 
\ln \Tr e^{-\beta H_n^{\rm Heis}}, \end{equation}
we compute again the cumulant generating function
\begin{eqnarray} C(t_1, t_2, t_3) &=& \ln \Tr e^{t_1 \sigma^x + t_2 \sigma^y + t_3 \sigma^z} \non \\ 
&=& \ln 2 \cosh \sqrt{t_1^2 + t_2^2 + t_3^2}. \end{eqnarray}
As in the previous example, we find that 
\begin{equation} I(x,y,z) = I_0(\sqrt{x^2 + y^2 + z^2}), \end{equation}
and therefore 
\begin{equation}  f_{\rm Heis}(\beta,J) = \inf_{\substack{(x,y,z) \in \RR^3: \\ x^2+y^2+z^2 \leq 1}}
\big\{-J(x^2 + y^2 + \Delta z^2) + \frac{1}{\beta} I_0(\sqrt{x^2+y^2+z^2}) \big\}. \end{equation}

\setcounter{equation}{0}

\section{General mean-field spin systems}

We now generalize  the above theorem to general symmetric functions of $q$ variables.
\begin{theorem}  Let $F:\RR \to \RR$ be a continuous function and let $Q$ be a symmetric polynomial. 
If $X_1,\dots,X_q, H \in {\cal M}$ are hermitian matrices then
\begin{eqnarray} \label{GMF}  \lefteqn{\lim_{n \to \infty} \frac{1}{n} \ln 
\Tr e^{n[F \circ Q(X_1^{(n)}, \dots, X_q^{(n)})]}} \non \\
&=& \sup_{(u_1,\dots,u_q) \in \prod_{i=1}^q {\rm co}(\sigma(X_i))} 
[F\circ Q (u_1,\dots,u_q) - I(u_1,\dots,u_q)], \end{eqnarray}  
where $I: \RR^q \to [0,+\infty]$ is the Legendre transform of
\begin{equation} \label{QuantumqGF} 
C(s_1,\dots,s_q) = \ln \Tr e^{s_1 X + \dots + s_q X_q}. \end{equation}
\end{theorem}

\Pf By Lemma~\ref{QLD-L4-4} we can approximate $F$ by a polynomial in which case $F \circ Q$ is 
also a symmetric polynomial, which we simply write as $Q$. 
Such a polynomial can be written as a linear combination of powers of linear combinations of the 
variables $x_1,\dots,x_q$ as follows.
\begin{equation} Q(x_1,\dots,x_q) = \sum_{r=1}^{M} \alpha_r Y_r(x_1,\dots,x_q)^{p_r}, \end{equation}
where $p_r  \leq  {\rm ord}(Q)$ and 
\begin{equation} Y_r(x_1,\dots,x_q) = \sum_{i=1}^q \zeta_{r,i} x_i. \end{equation}
It follows that 
\begin{eqnarray} \lefteqn{\lim_{n \to \infty} \frac{1}{n} \ln 
\Tr e^{n Q(X_1^{(n)}, \dots, X_q^{(n)})}} \non \\
&=& \sup_{(y_1,\dots,y_M) \in \RR^M} 
\big\{ \sum_{r=1}^M \alpha_r y_r^{p_r} - \tilde{I}(y_1,\dots,y_M) \big\}, \end{eqnarray}  
where 
\begin{equation} \tilde{I}(y_1,\dots,y_M) = \sup_{t_1,\dots, t_M \in \RR} 
\big\{\sum_{r=1}^M t_r y_r - \tilde{C}(t_1,\dots,t_M) \big\}, \end{equation}
and 
\begin{eqnarray} \tilde{C}(t_1,\dots,t_M) &=& \ln \Tr e^{\sum_{r=1}^M t_r Y_r(X_1,\dots,X_q)} 
\non \\ &=& \ln \Tr e^{\sum_{i=1}^q \left(\sum_{r=1}^M \zeta_{r,i} t_r \right) X_i}.
\end{eqnarray}
Now let
\begin{equation} \sum_{r=1}^M \zeta_{r,i} t_r = s_i, \mbox{ or } \ul{s} = \zeta^T \ul{t}, 
\mbox{ and hence }  \tilde{C}(\ul{t}) = C(\ul{s}). \end{equation}
We claim that $\tilde{I}(y_1,\dots,y_M) = +\infty$ unless there exist $u_1,\dots,u_q \in \RR$ such that  
$$ y_r = \sum_{i=1}^q \zeta_{r,i} u_i = \zeta\,\ul{u}. $$
Indeed, suppose that $\ul{y} \notin {\rm Ran}(\zeta)$, then write $\ul{y} = \zeta \ul{u} + \ul{z}$,
where $\ul{z} \perp {\rm Ran}(\zeta)$. Then we can write
\begin{eqnarray}  \tilde{I}(\ul{y}) &=& \sup_{\ul{t} \in \RR^M} \left[
\langle \ul{t}, \zeta \ul{u} + \ul{z} \rangle - C(\zeta^T \ul{t}) \right] \non \\ &=& 
\sup_{\ul{s} \in \RR^q} \sup_{\ul{t}' \perp {\rm Ran}(\zeta)} \left[
\langle \ul{t}',  \ul{z} \rangle + \langle \ul{s}, \ul{u} \rangle - C(\ul{s}) \right]. \end{eqnarray}
This equals $+\infty$ unless $\ul{z} = 0$. Inserting this into the above expression for 
$\lim_{n \to \infty} \frac{1}{n} \ln  \Tr e^{n[Q(X_1^{(n)}, \dots, X_q^{(n)})]}$, we obtain 
\begin{eqnarray} \lefteqn{\lim_{n \to \infty} \frac{1}{n} \ln 
\Tr e^{n Q(X_1^{(n)}, \dots, X_q^{(n)})}} \non \\
&=& \sup_{(u_1,\dots,u_q) \in \RR^q} 
\left[ \sum_{i=1}^q \alpha_r \left( \sum_{i=1}^q \zeta_{r,i} u_i \right)^{p_r}  - I(u_1,\dots,u_q)
\right] \non \\ &=& \sup_{(u_1,\dots,u_q) \in \RR^q} [Q(u_1,\dots,u_q) - I(u_1,\dots,u_q)]. \end{eqnarray}  

\textbf{Examples.}
\begin{enumerate} \item
An easy example is $Q(x_1,x_2) = x_1 x_2$. Clearly, 
$$ Q(x_1,x_2) = \frac{1}{4} [(x_1+x_2)^2 - (x_1-x_2)^2]. $$
The symmetrized version of $Q$ is $Q(X_1,X_2) = \half(X_1 X_2 + X_2 X_1)$.
\item
Similarly, if $Q(x_1,x_2) = x_1 x_2 x_3$. Then 
\begin{eqnarray*}  Q(x_1,x_2,x_3) &=& \frac{1}{24} \big( (x_1+x_2+x_3)^3 -  (x_1+x_2-x_3)^3 
\\ && \qquad - (x_1-x_2+x_3)^3 + (x_1-x_2-x_3)^3 \big). \end{eqnarray*}
In this case, therefore, we can take $M=4$, 
$$ \zeta = \left( \begin{array}{ccc} 1 & 1 & 1 \\ 1 & 1 & -1 \\ 1 & -1 & 1 \\ 1 & -1 & -1 \end{array} 
\right) $$ and $\alpha = \frac{1}{24} (1,-1,-1,1)$. 
There are various symmetrized versions of $Q$, for example $Q(X_1,X_2,X_3) = \half(X_1 X_2 X_3 + X_3 X_2 X_1)$, but also $Q(X_1,X_2, X_3) = \frac{1}{6} (X_1 X_2 X_3 + X_1 X_3 X_2 + X_2 X_1 X_3 + 
X_2 X_3 X_1 + X_3 X_1 X_2 + X_3 X_2 X_1)$. 
These are equivalent in the limit $n \to \infty$, namely, they can be replaced by 
$$ Q(\ul{X}_1,\dots, \ul{X}_n) = \frac{1}{n^3} \sum_{\substack{i_1, i_2. i_3 \in \{1,\dots,n\} \\ 
i_1 \neq i_2 \neq i_3 \neq i_1}} \frac{n!}{(n-3)!} X_{1,i_1} X_{2,i_2}  X_{3,i_3}.  $$
\item
Consider the more complicated example $Q(x_1,x_2) = x_1^3 x_2^3$. Then we obviously need 
the sixth power, so we compute
$$ (x_1 + x_2)^6 - (x_1 - x_2)^6 = 4 \big( 3x_1^5 + 10 x_1^3 x_2^3 + 3 x_1 x_2^4 \big). $$
This eliminates two terms. To eliminate the other two, we compute also 
$$ (2x_1 + x_2)^6 - (2x_1 - x_2)^6 = 8 \big( 48 x_1^5 + 40 x_1^3 x_2^3 + 3 x_1 x_2^4 \big). $$
By symmetry, it is obvious that we also need to compute
$$ (x_1 + 2x_2)^6 - (x_1 - 2x_2)^6 = 8 \big( 3 x_1^5 + 40 x_1^3 x_2^3 + 48 x_1 x_2^4 \big). $$
Adding these we have
\begin{align*} & (2x_1 + x_2)^6 - (2x_1 - x_2)^6 +  (x_1 + 2x_2)^6 - (x_1 - 2x_2)^6 \\
&\qquad = 8 \big( 51 x_1^5 + 80 x_1^3 x_2^3 + 51 x_1 x_2^4 \big). \end{align*}
It thus follows that
\begin{align*} & 34 \big( (x_1 + x_2)^6 - (x_1 - x_2)^6  \big)   
- (2x_1 + x_2)^6 + (2x_1 - x_2)^6 \\ & \qquad -  (x_1 + 2x_2)^6 + (x_1 - 2x_2)^6 \big) %\\ &\qquad 
= 720 x_1^3 x_2^3. \end{align*}

\end{enumerate}
\medskip

%\newpage

\setcounter{equation}{0}

\section{Appendix: Proof of the non-commutative H\"older inequality}

We need to generalize Hadamard's 3-line theorem:

\begin{lemma} \label{Hadamard} Consider the simplex 
$$ \Delta = \{(x_1,\dots,x_N) \in \RR^N:\, x_j\geq 0\, (j=1,\dots,N);\,x_1+\dots +x_N \leq 1\} $$
and the corresponding tubular set $\Delta \times \RR^N \subset \CC^N$. Suppose that 
$\phi: \Delta \times \RR^N \to \CC$ is bounded and continuous, and analytic in the interior. \\
If $|\phi(iy_1\dots,iy_N)| \leq M_0 > 0$ and $|\phi(iy_1,\dots,1+iy_k,\dots,iy_N)| \leq M_k > 0$
for $k=1,\dots,N$, and $y_1,\dots, y_N \in \RR$, then 
\begin{equation} \label{Hadamardineq} |\phi(z_1,\dots,z_N)| \leq M_0^{1-\Re(z_1) \dots - \Re(z_N)} 
\prod_{k=1}^N M_k^{\Re(z_k)}. \end{equation}
\end{lemma}

\Pf Replacing $\phi(z_1,\dots,z_N)$ by $\tilde{\phi} = \phi(z_1,\dots,z_N) M_0^{z_1 + \dots + z_N -1} 
\prod_{k=1}^N M_k^{-z_k}$ we can assume that $M_0 = M_1 = \dots = M_N = 1$. Indeed,
in that case $|\tilde{\phi}(iy_1,\dots,iy_N)| \leq 1$ and $|\tilde{\phi}(iy_1,\dots,1+y_k,\dots,iy_N)| \leq 1$
and if \\ $|\tilde{\phi}(z_1,\dots,z_N)| \leq 1$ then $\phi$ satisfies the bound (\ref{Hadamardineq}). 

Now if $\tilde{\phi}(z_1,\dots,z_N) \to 0$ if $|z_1| + \dots + |z_N| \to +\infty$ inside the tubular region
then it follows from the maximum modulus principle (see e.g. \cite{Vlad1966}, \S6.4) that 
$|\tilde{\phi}(z_1,\dots,z_N)| \leq 1$. Otherwise, consider the functions 
$$ \psi_n(z_1,\dots,z_N) = \tilde{\phi}(z_1,\dots,z_N) \prod_{k=1}^N e^{z_k^2/n} e^{-1/n}. $$
Since $\Re(z_k^2) = x_k^2 - y_k^2$ if $x_k = \Re(z_k)$ and $y_k = \Im(z_k)$ we have
$\sum_{k=1}^N \Re(z_k^2) \leq \sum_{k=1}^N x_k^2 \leq \left(\sum_{k=1}^N x_k \right)^2 \leq 1$
and therefore $|\psi_n(z_1,\dots,z_N)| \leq 1$, and also $\psi_n(z_1,\dots,z_N) \to 0$ as 
$|z_1| + \dots + |z_N| \to +\infty$. Therefore $|\psi_n(z_1,\dots,z_N)| \leq 1$, and taking $n \to \infty$
it follows that $\tilde{\phi}(z_1,\dots,z_N)| \leq 1$. \qed

\textbf{Proof of Lemma~\ref{QLD-L3-3} }
Let $A_k = U_k |A_k|$ ($k=1,\dots,N$) be the polar decompositions. 
We apply Lemma~\ref{Hadamard} to the function
$$ F(z_1,\dots,z_{N-1}) = \Tr \left(\prod_{k=1}^{N-1} \big(U_k |A_k|^{p_k z_k} \big)
U_N |A_N|^{p_N(1-(z_1+\dots+z_{N-1}))} \right). $$
Then, for $y_1,\dots,y_{N-1} \in \RR$,
\begin{eqnarray*} \lefteqn{|F(iy_1,\dots,iy_{N-1})|} \\ &=& 
\left|\Tr \left(\prod_{k=1}^{N-1} \big(U_k |A_k|^{i y_k p_k} \big)
U_N |A_N|^{p_N}\,|A_N|^{-i p_N(y_1+\dots+y_{N-1})} \right) \right| \\ 
&\leq & \Tr (|A_N|^{p_N}) = ||A_N||_{p_N}^{p_N}. \end{eqnarray*}
and for $l=1,\dots,N-1$, 
\begin{align*} &|F(iy_1,\dots,1+iy_l, \dots,iy_{N-1})| \\ &= 
\left|\Tr \left(\prod_{k=1}^{l-1} \big(U_k |A_k|^{i y_k p_k} \big) 
U_l |A_l|^{p_l} |A_l|^{iy_l p_l} \right. \right.
\\ &\qquad \times  \left. \left.
\prod_{k=l+1}^{N-1} \big(U_k |A_k|^{i y_k p} \big) 
U_N \,|A_N|^{-i p_N(y_1+\dots+y_{N-1})} \right) \right| \\ 
&\leq  \Tr (|A_l|^{p_l}) = ||A_l||_{p_l}^{p_l}. \end{align*}
By Lemma~\ref{QLD-L3-1} therefore,  $$  |F(z_1,\dots,z_{N-1})| 
\leq \prod_{k=1}^{N-1} ||A_k||_{p_k}^{p_k x_k} 
||A_N||_{p_N}^{p_N x_N}, $$ where $x_N = 1-(x_1+\dots+x_{N-1})$.
Setting $x_k=1/p_k$, the result follows. \qed
\medskip

\textbf{Note.} The author confirms that he has no conflicting interests.

\end{document}